\documentclass[a4paper,fleqn,usenatbib,useAMS]{mnras}

\usepackage{mathptmx}

\usepackage[T1]{fontenc}
\usepackage{ae,aecompl}

\usepackage{etoolbox}
 
\usepackage{graphicx}	
\usepackage{amsmath}	
\usepackage{amssymb}	
\usepackage{multicol}   

\newcommand{\kms}{\,km\,s$^{-1}$} 
\newcommand{\aud}{\,AU\,d$^{-1}$} 
\newcommand{\ms}{\,m\,s$^{-1}$} 
\newcommand{\water}{H$_{2}$O} 
\newcommand{\meth}{CH$_{3}$OH} 
\newcommand{\ngci}{NGC~6334I} 
\newcommand{\ngcimm}{NGC~6334I-MM1} 
\newcommand{\Stwofivefive}{S255IR-NIRS3} %
\newcommand{\Gthreetwothree}{G323.46$-$0.08} %
\newcommand{\Gthreefiveeight}{G358.93$-$0.03} %

\makeatletter\ifdefined\Hy@StartlinkName
\def\addOneNestingLevelStartLink{%
  \gdef\Hy@StartlinkName##1##2{%
    \sbox0{\Hy@StartlinkNameOrig{##1}{##2}}\usebox0
    \global\let\Hy@StartlinkName\Hy@StartlinkNameOrig%
  }%
}
\def\addOneNestingLevelEndLink{%
  \gdef\pdfendlink{%
    \sbox0{\pdfendlinkOrig}\usebox0%
    \global\let\pdfendlink\pdfendlinkOrig%
  }%
}
\let\Hy@StartlinkNameOrig\Hy@StartlinkName
\let\pdfendlinkOrig\pdfendlink
\else
\let\addOneNestingLevelStartLink\relax
\let\addOneNestingLevelEndLink\relax
\fi

\title[Ten years since the MM1 accretion outburst]{Ten years since the \ngcimm\ accretion 
outburst: the impact on its environment}

\author[G. MacLeod et al.]{G. C. MacLeod$^{1,2,3,4}$\thanks{E-mail: MacLeod@gerafoundation.com}, 
A. Caratti o Garatti$^{5}$, W. A. Baan$^{1,6}$, D. P. Smits$^{7}$, 
\newauthor{T. R. Hunter$^{8}$, C. L. Brogan$^{8,9}$, J. O. Chibueze$^{7, 10}$, R. A. 
Burns$^{11}$, B. Stecklum$^{12}$, }
\newauthor{V. Wolf$^{12}$, J. M. Vorster$^{13}$, A. Jengis$^{1,2}$,
J. Esimbek$^{1,2}$, and J. Quick$^{4}$}\\ 
\\
$^{1}$State Key Laboratory of Radio Astronomy and Technology, Xinjiang Astronomical
Observatory, CAS, 150 Science 1-Street,\\ Urumqi, Xinjiang 830011,  China \\
$^{2}$University of Chinese Academy of Sciences, Beijing 100080, China \\
$^{3}$The Open University of Tanzania, P.O. Box 23409, Dar-Es-Salaam, Tanzania \\
$^{4}$Hartebeesthoek Radio Astronomy Observatory, PO Box 443, Krugersdorp, 1741, South Africa \\
$^{5}$Istituto Nazionale di Astrofisica Osservatorio Astronomico di Capodimonte Napoli 
Salita Moiariello, 16, 80131 - Napoli, Italy\\
$^{6}$Netherlands Institute for Radio Astronomy, ASTRON, 7991 PD, Dwingeloo, The Netherlands\\
$^{7}$UNISA Centre for Astrophysics and Space Sciences (UCASS),
College of Science, Engineering and Technology, \\ University of South Africa, Cnr Christian de Wet Rd and Pioneer Avenue, Florida Park, 1709, Roodepoort, South Africa.\\
$^{8}$National Radio Astronomy Observatory, 520 Edgemont Rd, Charlottesville, VA 22903, USA\\
$^{9}$Department of Astronomy, University of Virginia, P.O. Box 3818, Charlottesville, 
VA 22904, USA\\
$^{10}$Department of Physics and Astronomy, Faculty of Physical Sciences, University of 
Nigeria, Carver Building,\\ 1 University Road, Nsukka, 410001 Nigeria\\
$^{11}$Mizusawa VLBI Observatory, National Astronomical Observatory of Japan, 2-21-1 
Osawa, Mitaka, Tokyo 181-8588, Japan \\
$^{12}$Th\"uringer Landessternwarte, Sternwarte 5, 07778 Tautenburg, Germany\\
$^{13}$ Department of Physics, P.O. box 64, FI- 00014, University of Helsinki, Finland \\ 
}

\date{Accepted ???. Received ???; in original form 2026 June 3}

\pubyear{2026}

\begin{document}
\label{firstpage}
\pagerange{\pageref{firstpage}--\pageref{lastpage}}
\maketitle

\begin{abstract}
Ten years on, flaring continues in masers associated with \ngcimm. Methanol and 
hydroxyl masers are nearing quiescent values, but water features remain elevated. 
Several more 1.665\,GHz hydroxyl masers, notably, that flared during the accretion 
event in \ngcimm, are identified and presented. Some hydroxyl masers appear to have 
been active prior to the 2015 event. Possible explanations are discussed. The radial 
infrared expansion speed during the accretion event is estimated to be $\sim$1.8\aud. 
However, the expansion speed along the North-South outflow is about ten times 
faster ($\sim$18\aud). The maser flaring activity illuminated clumpy structure in 
\ngcimm. Quasi-periodicity is reported after the 2015 accretion event in the total 
integrated 12.2\,GHz methanol emission, predominantly in $v_{12.2}=[-12.0,-9.9]$
\kms, after the accretion event in \ngcimm. It is not certain if the periodic masers 
are located in \ngcimm, -MM2, or -MM3. The magnetic field, measured from the Zeeman 
splitting of two ground state hydroxyl lines, undergoes changes before and after the 
flare.

\end{abstract}

\section{Introduction}
\citet{mgk93} suggested that Class II methanol (\meth) masers are not significantly 
variable (<25$\%$); were they wrong! Three years later \citet{mg96} discovered 
significant variability in 1.665\,GHz hydroxyl (OH) and 6.7\,GHz \meth\ masers 
associated with G351.78--0.54. These results prompted the launch of a multi-source 
long-term monitoring programme at the Hartebeesthoek Radio Astronomy Observatory 
(HartRAO). \citet{ggv03, glgw09} made the observing in this programme more evenly 
spaced and increased cadence, resulting in the discovery of periodic \meth\ masers. 
Several other monitoring programmes globally reported many more \citep{Aetal10,setal11,
fetal14,metal15,metal16,swb15,Setal17, Szetal18, Proven19,Olech2020,Olech2022,Tanabe2023,
Szmczak2024,Wolak2025}. Clearly, class II \meth\ masers are significantly variable.

It is not surprising that masers associated with massive star-forming regions (MSFRs) 
are variable. Anything that causes variations in the cloud conditions in which the 
masers reside, or seed photon production, or pumping mechanisms will cause flaring. 
Generally, water (\water) masers are believed to be collisionally pumped and hence 
vary significantly during ejection bursts and/or outflows/jets causing shock fronts 
\citep{Elitzur1989}. OH and Class II \meth\ masers, e.g.\ 6.7 $\&$ 12.2\,GHz, 
associated with MSFRs are suspected to be radiatively pumped. Periodic masers may be 
explained by five scenarios: (1) proto-stellar pulsations \citep{Ietal13}, (2) rotating 
spiral shocks \citep{ps14}, (3) modulated accretion in a binary system \citep{Aetal10}, 
(4) interacting winds of a proto-binary star system \citep{vdwetal09,vdwetal16}, or (5) 
superradiance \citep{RHM23,HRM24,RAB25,RAP26}.

Another common cause of maser flaring in MSFRs is the occurrence of accretion bursts 
\citep{Moscadelli17}. In 2015 two major accretion outbursts were detected in MSFRs: 
\Stwofivefive\ \citep{Caratti17} and \ngcimm\ \citep{hunter17b}. Both outbursts 
experienced maser flaring, reported by \citet{Fujisawa2015} and \citet{Szetal18} 
in \Stwofivefive, and \citet{MacLeod2018} in \ngcimm\ triggered by the accretion bursts. 

These discoveries and the associated 6.7\,GHz \meth\ maser flaring contributed to the 
creation of the Maser Monitoring Organization (M2O)\footnote{Visit the M2O website at 
MaserMonitoring.org for more information.}. One of the primary functions of the M2O is 
the early identification of accretion bursts in massive young stellar objects (MYSOs).
Indeed, in 2019 the M2O was notified of rapidly strengthening 6.7\,GHz \meth\ masers in 
\Gthreefiveeight\ \citep{Sugiyama2019}. The M2O initiated multi-wavelength observations 
globally, resulting in the detection of flaring in many maser transitions \citep{Breen2019,
MacLeod2019,Volvach2020,Miao2022}. The flaring was attributed to a ``heat" wave that 
propagated radially outward at $\geq$4 per cent the speed of light \citep{Burns20}. 
Infrared observations reported by \citet{Stecklum2021} confirmed that it was also a major 
accretion outburst. 

A fourth accretion outburst was proposed by \citet{Proven19,MacLeod2021B}, from its 6.7\,GHz \meth\ masers in \Gthreetwothree\ 
and confirmed by \citet{Wolf2024} in the infrared. \citet{Proven19} also reported the 
6.7\,GHz\ \meth\ masers in \Gthreetwothree\ are periodic ($P\sim93$\,d). \citet{Wolf2024} 
proposed that an accretion burst can induce periodicity.

The impact of accretion events on MSFRs and masers is significant. Multi-epoch maser spot 
maps produced during the accretion event in \Gthreefiveeight\ \citep{Burns20} illuminated 
a possible spiral structure. They further show the disruption, thermalisation, and/or 
destruction of \meth\ masers as the ``heat" wave produced by the released energy expanded 
outward. Detailed source structure could not be identified in the other accreting sources 
due to the lack of multi-epoch interferometric radio frequency observations.

The nomenclature we have adopted here for the source name is \ngci\ \citep{DeBuizer02}. It 
is comprised of seven millimetric hot-cores we refer to as MM1-7 \citep{Brogan2016}. Each 
hot-core is further segregated into maser clusters \citep{hunter18} which we denote by 
capital letters. For example the accretion event occurred in the hot-core \ngcimm\ 
\citep{hunter17b}; we refer to this region as MM1. The accretion event is centred on the 
\meth\ maser cluster \ngcimm B \citep{hunter18}; we refer to this cluster as MM1B. Also 
\meth\ masers were originally only identified in MM2 and MM3 \citep{elletal96}; we refer to 
these regions as MM2/3. Names of other sources are either Galactic positions or as they 
appear in their reference.

\ngci, at $d$ = 1.3 kpc \citep{chibueze14}, also has multiple outflows centred at MM1B
where the accretion burst occurred. This leads to a complex envelope geometry 
\citep{Hunter2021,Chibueze2021,Vorster2024}. Multi-epoch JVLA observations have shown 
that the \meth\ masers in MM1 turned on for the first time after the accretion burst 
\citep{hunter18}. This can be explained by a thermal ``heat" wave propagating up to 
10\,000 AU, traced by thermal \meth\ lines \citep{Wu2023}. The infrared radiation in 
turn desorbs \meth\ ice off of dust grains, and provides the necessary infrared pumping 
($T_{\rm dust} \sim 400$ K) for bright 6.7 GHz \meth\ masers \citep{Guadarrama2024}. 
However, the 22\,GHz \water\ masers in MM1 behaved strangely because those masers 
closest to the bursting source disappeared, while the masers at a bow shock, identified 
as CM2 in \citet{Brogan2018}, 2700\,AU north of MM1B, flared by orders of magnitude. 
The disappearance of these masers can be attributed to the collisional pumping of 22\,GHz 
\water\ masers, while the flaring in CM2 requires radiative heating of the collisional 
partner H$_2$ \citep{g12, 1995CAS....28.....W}. The complementary nature of various maser 
species (e.g.\ collisionally pumped 22\,GHz H$_2$O versus radiatively pumped 6.7\,GHz 
\meth) allows a multi-faceted view of the time evolution of the bursting source. 

Masers continue to flare in MM1 ten years after their onset. In this work, we focus on the 
maser transitions monitored at HartRAO through the 2015 accretion event in MM1: 1.665\,GHz 
OH, 6.7 and 12.2\,GHz \meth, and 22.2\,GHz \water. We identify flaring OH maser features 
not identified in \citet{MacLeod2018}, and analyse the continuing flaring of all these masers. 
We present possible evidence of the impact of the ongoing accretion outburst in \ngci\ on its 
surrounding environment.

\section{Observations}
Observations centred on the brightest \meth\ maser in MM3 at R.A.~=~$17^{h}~20^{m}~53\fs4$ 
and Dec.~=~$-35\degr$~47$\arcmin$~1$\farcs$5 (J2000) continue as part of the maser monitoring 
programme at HartRAO\footnote{See http://www.hartrao.ac.za/old.php for further information.} 
\citep{ggv03}. The observing methodology employed is described in \citet{MacLeod2018,MacLeod21}. 
Basic information for each transition is included in Table \ref{tab:transitions}. The 3$\sigma$ 
root-mean-square (rms) noise level achieved in each observation is $\sim 1.0$\,Jy. All 
observations are made in right and left circular polarisation (RCP and LCP). The 6.7 and 
12.2\,GHz \meth\ masers have been monitored for over 25 yrs, the 22.2\,GHz \water\ masers for 
12 yrs, and the 1.665\,GHz OH masers for 13 yrs.

\begin{table}
\caption{Parameters of the HartRAO receiver beam sizes and velocity resolution employed.}
\label{tab:transitions}
\begin{tabular}{ccccc}
Wavelength  &Frequency  &Beam size  &Velocity resolution \\
cm          &GHz        &arcmin     &\kms                 \\ \hline
1.3         &22.2       &2.2        &0.105 \\
2.5         &12.2       &4.0        &0.048 \\
4.5         &6.7        &7.0        &0.044 \\
18          &1.7        &29.6       &0.044 \\ \hline
\end{tabular}
\end{table}

\section{Data Analysis}
In this work we select and analyse velocity features consisting of a simple extent 
of sequential velocity channels in a spectrum. It is important to note that each 
velocity feature may not always be a single maser, as they can be possibly line-blended 
masers in a narrow velocity extent. Nonetheless, we are able to extract information 
by analysing velocity features. The line-blended masers comprising a velocity feature 
may be spatially coincident and/or affected similarly. Velocity features that are much 
brighter than, and/or isolated from, other possible line-blending features offer the 
best targets for study. 

The first method of analysis is to select a single velocity channel in the spectrum 
central to the velocity feature and produce a time series of the channel flux densities, 
$F_\textrm{C}$, from this. A second method is to fit Gaussian profiles to the velocity 
feature in each epoch of observation and determine the flux density, $F_\textrm{G}$, 
the line width, $w$, and the velocity, $v$. Highly accurate velocities and line widths 
(of the order of m\,s$^{-1}$) can be determined. Creating automated methods to fit these 
Gaussians have limitations, which are minimised through deletion of outliers from the 
trends visible in the time series. 

Time series of integrated flux density in a particular velocity range may also be 
produced, providing a general overview of the impact of an accretion event on the 
entire affected region. Such time series have been created for MM1 in the velocity 
range $v = [-9.0, -3.0]$\kms\ and for MM2 and/or MM3 (MM2/3) over $v = [-12.0, -9.9]$\kms\ 
for dominant 6.7 and 12.2\,GHz maser emission. Hereafter, we refer to these velocity 
ranges as $V_\textrm{MM1}$ and $V_\textrm{MM2/3}$, respectively.

\section{Results: The Flaring continues}
\subsection{Methanol and water masers}
The 6.7\,GHz \meth\ masers associated with \ngci\ have been monitored from 1999 to 
the present. Evidence of two distinct flaring events in MM1 are present in the data: 
in 1999 \citep{ggv05} and 2015 \citep{MacLeod2018}. These two events make it possible 
to study maser activity leading up to, during, and after the 2015 accretion event. 
This study is augmented by observations of 12.2\,GHz \meth\ (from 2000 onward), 
1.665\,GHz OH (from 2011 onward), and 22.2\,GHz \water\ (from 2013 onward) masers. 
Initial results of monitoring these four masering transitions in MM1 through the 
2015 flaring event, are reported in \citet{MacLeod2018}. 

\begin{figure}   
	\includegraphics[clip,width=\columnwidth]{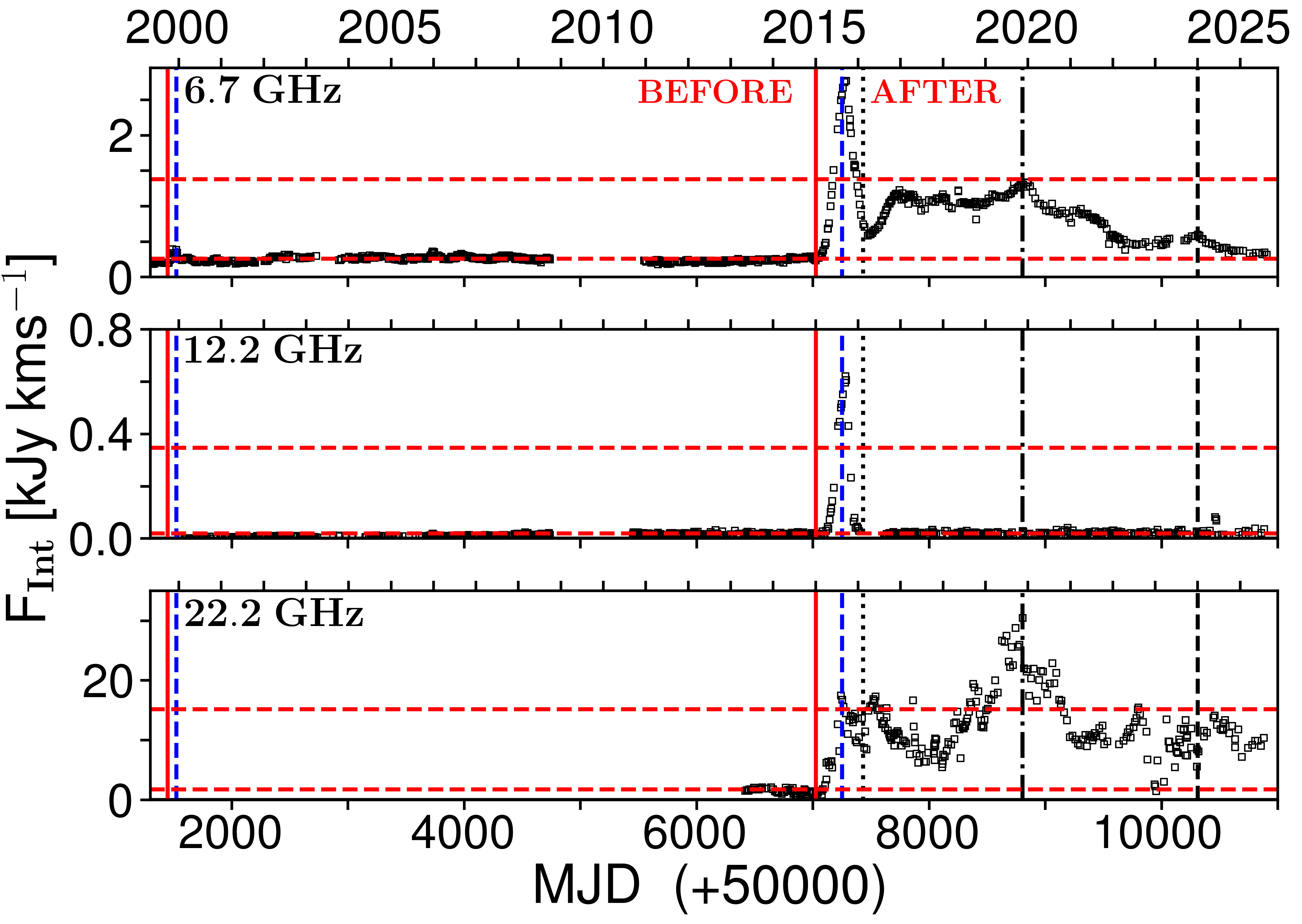}
	\caption{Integrated flux density time series plots for 6.7 $\&$ 12.2\,GHz  \meth\ 
    and 22.2\,GHz \water\ maser emission in $V_\textrm{MM1}$. Solid vertical red lines 
    demarcate the onset of the two flares, on MJD 51446 (1999 September 25) and MJD 57023 
    (2015 January 1). The 6.7 and 12.2\,GHz \meth\ masers reached their maxima, indicated 
    by blue dashed lines, on MJD 51522 (1999 December 9) and 57249 (2015 August 15), 
    respectively. The brightest OH maser reached its maximum on MJD 57429 (2016 February 
    11), and is identified by the dotted vertical black line. Water masers reached their 
    maximum (the dot-dashed vertical black line) on MJD 58803 (2019 November 16). The 
    dashed vertical black line indicates a secondary flare in the 6.7\,GHz \meth\ masers 
    that peaked on MJD 60310 (2024 January 1). The dashed horizontal lines demarcate the 
    quiescent and half-power flux densities for each transition.}
 \label{fig:IntfluxMeth}
\end{figure}

The time series of the integrated flux density of the 6.7 and 12.2\,GHz \meth\ 
emission in $V_\textrm{MM1}$ are plotted in Fig.\ \ref{fig:IntfluxMeth}. The 
integrated flux density of the 22.2\,GHz \water\ maser emission in its entire 
velocity extent, $v = [-61, +47]$\kms, is also plotted in this image. There is 
a stark difference between the energy of the 6.7\,GHz flare and the time scale 
of its recovery in 2015 compared to those characteristics in the much less intense 
1999 flare event \citep{MacLeod2018}, as can be seen in Fig.\ \ref{fig:IntfluxMeth}. 

Observations of the 12.2\,GHz \meth\ masers presented in Fig.\ \ref{fig:IntfluxMeth} 
began four months after cessation of flaring activity in the 1999 event in MM1; they 
showed no evidence of flaring until the 2015 event. After the MM1 flaring in 2015, the 
12.2\,GHz \meth\ masers returned to quiescence within 13 months of the flare onset. 
In comparison, the 6.7\,GHz emission remains active; it decayed from its maximum, to 
below its half-power value, and then underwent secondary flaring, increasing back to 
its half-power value at the same time that the 22.2\,GHz \water\ emission reached its 
maximum. Both the 6.7 and 22.2\,GHz maser transitions were decaying towards quiescent 
values, however, recently they reached a new maxima on MJD 60310 (2024 January 1). 
Ten years on from the 2015 outburst, activity continues in the 6.7\,GHz \meth\ and 
22.2\,GHz \water\ masers.

\begin{figure}   
	\includegraphics[clip,width=\columnwidth]{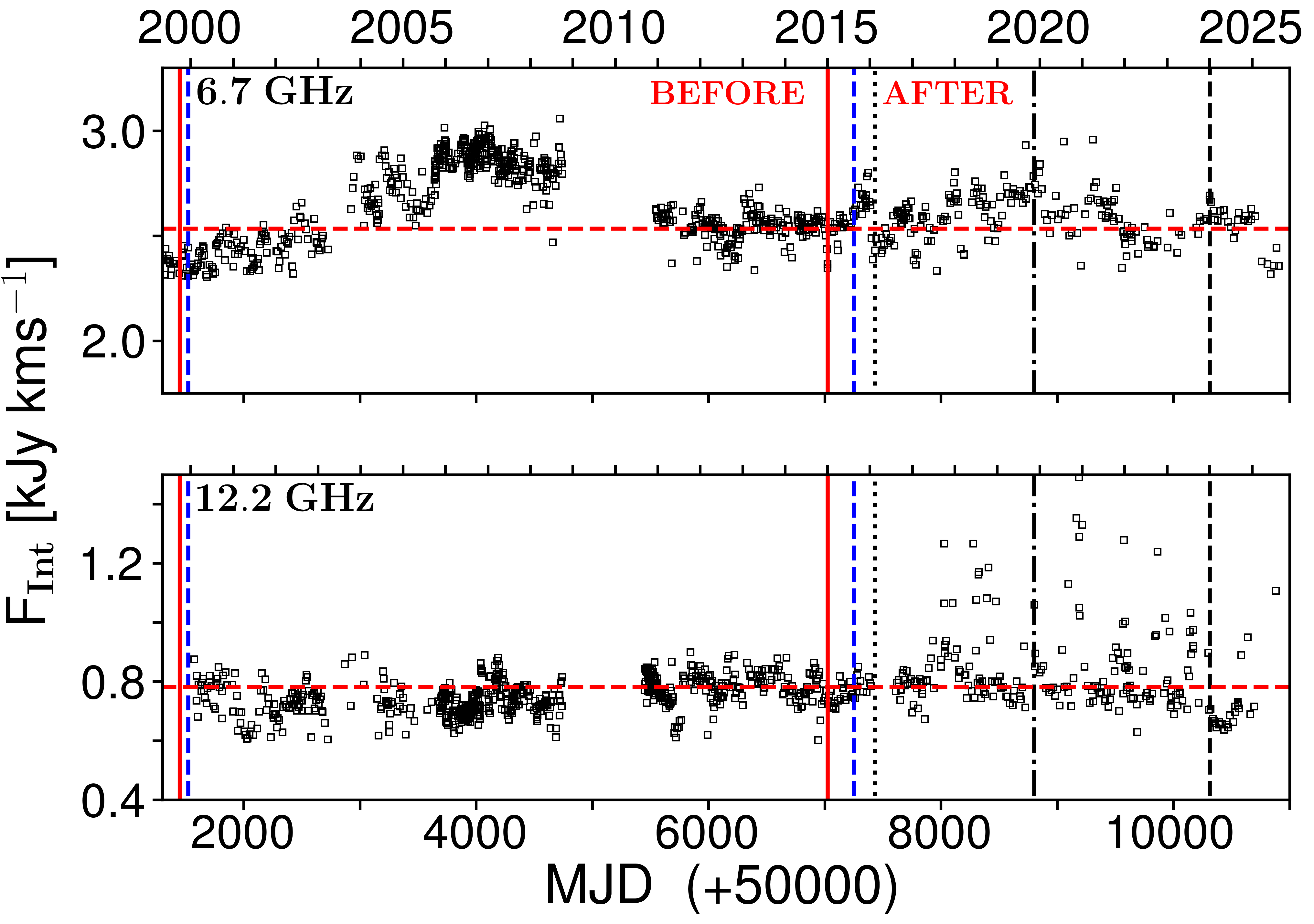}
	\caption{Integrated flux density time series plots for 6.7 $\&$ 12.2\,GHz \meth\ in 
    $V_\textrm{MM2/3}$. The vertical lines are as defined in Fig.\ \ref{fig:IntfluxMeth}. 
    The dashed horizontal lines denote the average values before MM1 flared.}
 \label{fig:IntfluxMethMM3}
\end{figure}

In Fig.\ \ref{fig:IntfluxMethMM3} the time series for the integrated flux densities 
in $V_\textrm{MM2/3}$ are plotted for both 6.7 and 12.2\,GHz \meth\ maser emission. 
The 6.7\,GHz \meth\ masers were slowly increasing from 2000 to 2008 with no associated 
12.2\,GHz flux density increase. Weak 6.7\,GHz \meth\ maser emission flaring, associated 
with the 2015 accretion event, is visible. On the other hand, the 12.2\,GHz \meth\ 
maser begins flaring about 800\,d (in 2017) after the 2015 MM1 event. The dispersion of 
the integrated flux density is about $\pm60$\,Jy\,\kms. New 12.2\,GHz flaring beginning 
in 2017 reaches maxima $\sim400$\,Jy\,\kms\ above the average value of 800\,Jy\,\kms.

\begin{figure}   
	\includegraphics[clip,width=\columnwidth]{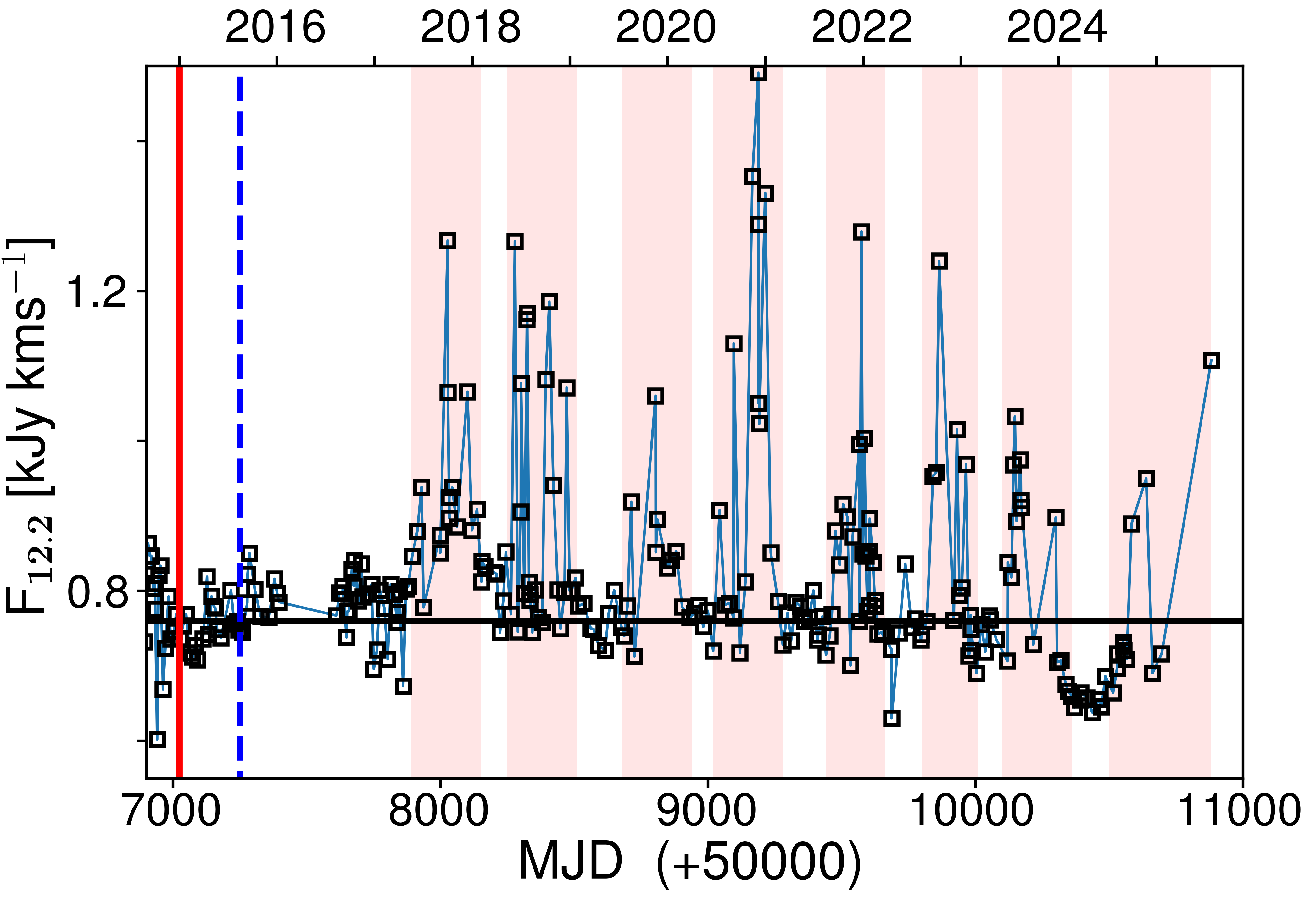}
	\caption{Time series plot of the integrated flux density of the 12.2\,GHz \meth\ 
    masers in $V_\textrm{MM2/3}$. Shaded red regions denote areas of elevated maser 
    activity. The vertical red and dashed blue lines denote the onset of MM1 flaring and 
    the peak of the $v_{6.7} = -5.24$\kms\ maser, respectively. The horizontal solid 
    black line marks the average integrated flux density prior to the flaring.}
 \label{fig:12blowup}
\end{figure}

We plot an abbreviated MJD extent of the 12.2\,GHz emission in $V_\textrm{MM2/3}$ in 
Fig.\ \ref{fig:12blowup} and highlight areas of increased maser activity. The eight 
flaring events are included in Table \ref{tab:Maxima} where the quoted errors represent 
the MJD range of elevated maser emission in each flare. The average separation of these 
peaks is $P_\textrm{Visible} = 380\pm50$\,d suggesting quasi-periodic activity. The 
duration of elevated maser emission occurs in an average MJD extent of $130\pm30$\,d, 
or about a third of the flare cycle.

\begin{table}
\centering
\caption{Determination of MJD dates of onset and maxima of the various time series 
reported here. The onset of the accretion event is estimated to be MJD 57023 (or 
2015 January 1) when the \meth\ and \water\ masers began flaring. Flare peaks grouped
in specific MJD extents, referenced as Regions I, II, III, and IV (see Fig.\
\ref{fig:Max_Dates}), are identified in boldface black, blue,  
red, and green colours respectively.}
\label{tab:Maxima}
\setlength{\tabcolsep}{3pt} 
\begin{tabular}{llcll}
\hline
Mol. & Freq.  & Pol. & Velocity & MJD peaks \\
 & (GHz) & & (\kms) & (+50000) \\
\hline 
OH & 1.665 & LCP & --7.26$^{1}$ & 7122, 8275\\
 &  &  & --7.26$^{2}$ & \textbf{\textcolor{black}{7167}}, \textbf{\textcolor{red}{7316}}, 8275\\
 &  &  & --7.53$^{1}$ & \textbf{\textcolor{red}{7347}}\\
 &  &  & --8.10$^{1}$ & 7429\\
 &  &  & --8.89$^{2}$ & \textbf{\textcolor{black}{7150$^{3}$}}\\
 &  &  & --9.46$^{1}$ & \textbf{\textcolor{red}{7356}}, 7786, 8130, 8446\\
 &  & RCP & --7.35$^{1}$ & \textbf{\textcolor{red}{7347}}, 7680\\
 &  &  & --7.79$^{1}$ & \textbf{\textcolor{red}{7332}}, 7528, 8097\\
 &  &  & --8.58$^{1}$ & \textbf{\textcolor{red}{7316}}, 7870\\
 &  &  & --8.89$^{1}$ & \textbf{\textcolor{red}{7348}}, 7870, 7991, 8432, \\
 &  &  &  &  8585\\
 &  &  & --11.86$^{2}$ & \textbf{\textcolor{black}{7123}}\\
\meth\	& 6.7 & I & [--9,--3]$^{4}$ & 1489$^{5}$, \textbf{\textcolor{blue}{7286}}, \textbf{\textcolor{red}{7356}}, \textbf{\textcolor{green}{7744}}, \\
 &  &  &  & 8118, 8543, 8729, 9306,\\
 &  &  &  & 10310\\
 &  &  & [--12,--9.9]$^{4}$ & 7374, 8118\\
&  &  & --6.74$^{1}$ & \textbf{\textcolor{black}{7151}}\\
 & 12.2 & I & [--9,--3]$^{4}$ & \textbf{\textcolor{blue}{7279}}\\
 &  &  & [--12,--9.9]$^{4}$ & 8020$\pm$130, 8380$\pm$130, \\
 &  &  & & 8810$\pm$130, 9150$\pm$130,\\
 &  &  & & 9550$\pm$110, 9905$\pm$105, \\
 &  &  & & 10230$\pm$130, 10690$\pm$190\\
\water\ & 22.2 & I & [--9,--3]$^{4}$ & \textbf{\textcolor{blue}{7242}}, 7539, 8378, 8803,\\
 &  &  &  &  9061, 9800\\
\hline
\multicolumn{5}{l}{$^{1}$Results for the velocity channel time series.}\\
\multicolumn{5}{l}{$^{2}$Results for the Gaussian fitted time series.}\\
\multicolumn{5}{l}{$^{3}$Value is the average over a range of epochs - due to bad data.}\\
\multicolumn{5}{l}{$^{4}$Results for the integrated flux density time series }\\
\multicolumn{5}{l}{~in the listed velocity extent.}\\
\multicolumn{5}{l}{$^{5}$Earlier methanol maser flaring event peak in 1999.}\\
\end{tabular}
\end{table}

\subsection{Hydroxyl masers}
\label{ohMM1}
\subsubsection{Flaring OH masers}
\begin{figure}   
	\includegraphics[clip,width=\columnwidth]{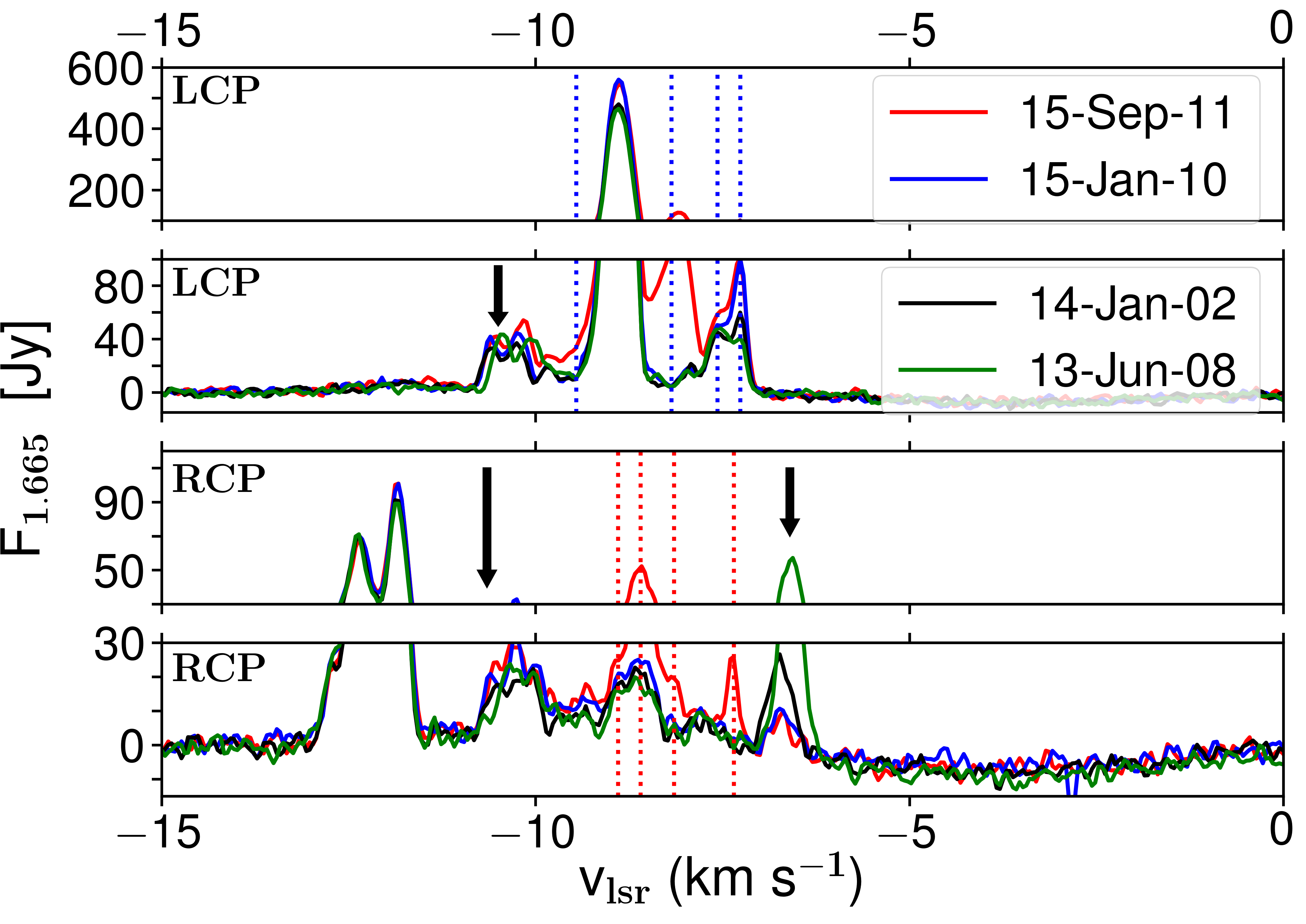}
	\caption{Multiple spectra of RCP (bottom two panels) and LCP (top two panels) 
    1.665\,GHz OH masers associated with \ngci\ at different epochs are plotted. 
    The vertical dashed lines denote specific velocity features to be analysed here, 
    red and blue are RCP and LCP features, respectively. Black arrows denote the 
    undulating, and hence contaminating, velocity features  associated with G351.161+0.697.}
 \label{fig:spectra}
\end{figure}

\begin{figure*}   
	\includegraphics[clip,width=\textwidth]{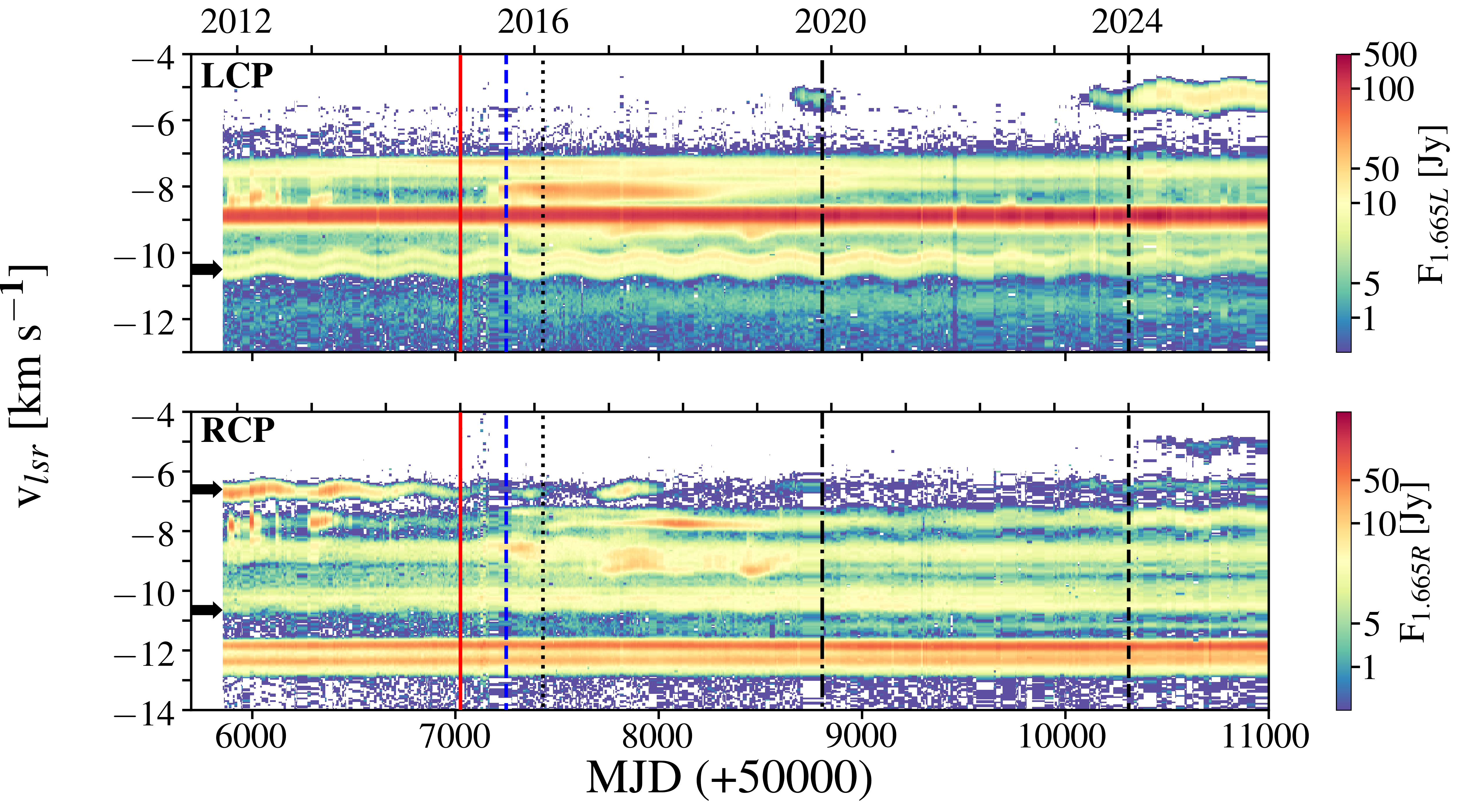}
	\caption{Dynamic spectra of 1.665\,GHz RCP and LCP OH masers associated 
    with MM3 are presented. The vertical lines are as defined in Fig.\ 
    \ref{fig:IntfluxMeth}. Black arrows denote the annually-undulating, and hence 
    contaminant (see \S~\ref{ohMM1}), velocity features associated with G351.161+0.697.}
 \label{fig:ds_oh1665}
\end{figure*}

Several OH spectra of 1.665\,GHz RCP ($v_\textrm{1.665R}$) and LCP ($v_\textrm{1.665L}$) 
are plotted in Fig.\ \ref{fig:spectra}. Flaring associated with the MM1 2015 event is 
visible, in particular in the feature $v_\textrm{1.665L} = -8.10$\kms, and reported in 
\citet{MacLeod2018}. Others are present but are easier to identify in the dynamic spectra 
plotted in Fig.\ \ref{fig:ds_oh1665} for both 1.665\,GHz OH circular polarisations. 

Line blending with features associated with MM3 and G351.161+0.697 \citep[1.665\,GHz 
velocity extent of --13.15 to --3.93\kms\ from MM3,][]{arm00} complicate identification. 
Those of G351.161+0.697 are more easily identified because they are $\sim$16\arcmin\ 
off-centre, in the 29$\farcm$6 HartRAO 18\,cm beam. As a result an incorrect velocity 
correction is applied inducing an annual periodic velocity drift ($\sim \pm$100\ms), 
e.g.\ features $v_\textrm{1.665L} \sim -10.2$ and --10.6\kms, visible in Fig.\ 
\ref{fig:ds_oh1665}. The two LCP examples given are sufficiently separated in velocity 
to have no impact on feature selection here. However, it is particularly problematic 
in our RCP observations shown in Fig.\ \ref{fig:ds_oh1665} where velocity feature 
overlap between the sources is greatest. Also, flux density variations in the MM3 OH 
masers add to this confusion. Fortunately, the 2015 MM1 accretion event can be used as 
a discriminator via the timing of their flaring. 

\begin{figure}   
	\includegraphics[clip,width=\columnwidth]{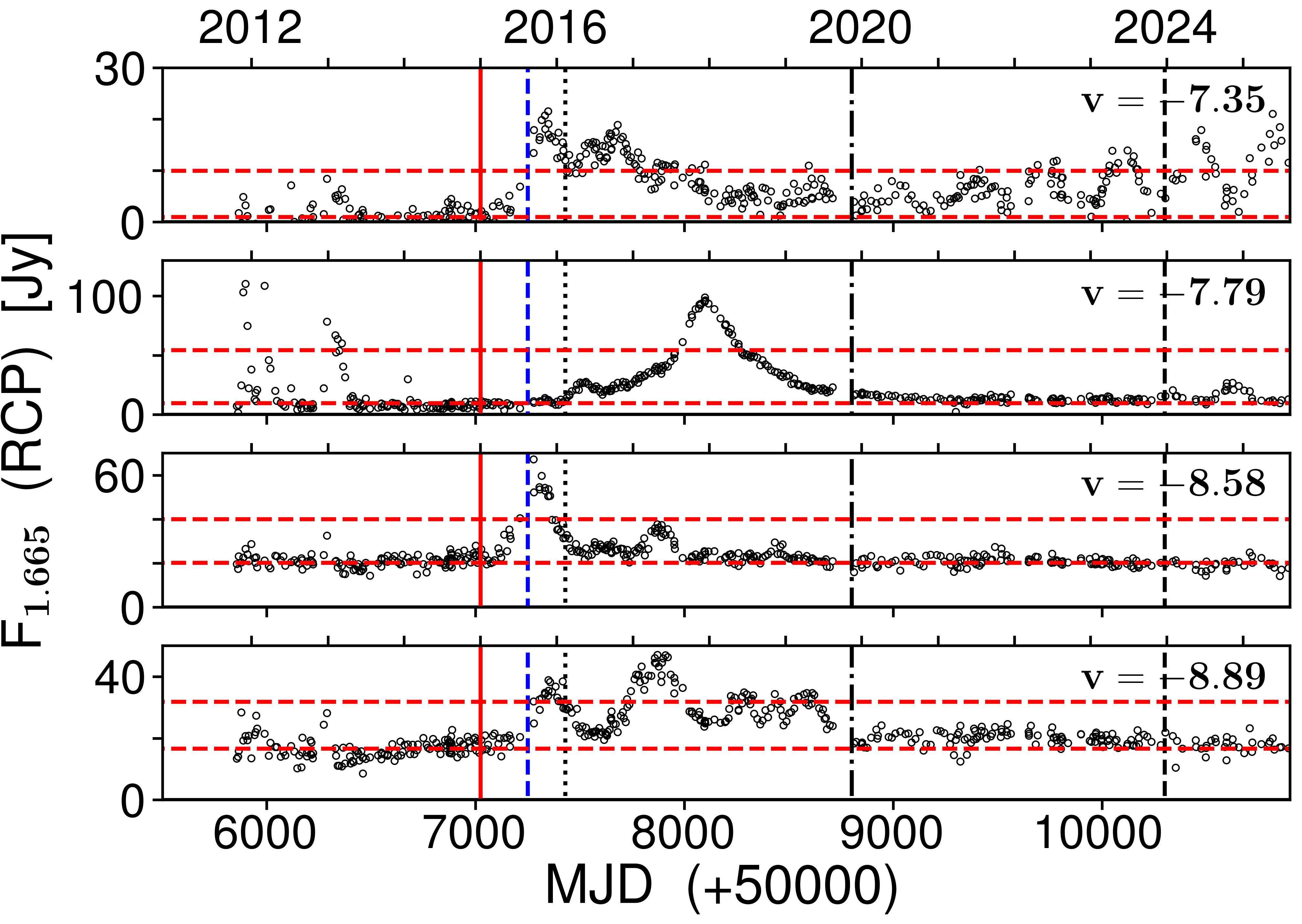}
	\caption{Time series plots of four 1.665\,GHz RCP OH velocity features are 
    presented. The vertical lines are as defined in Fig.\ \ref{fig:IntfluxMeth}.}
 \label{fig:RCP}
\end{figure}

\begin{figure}   
	\includegraphics[clip,width=\columnwidth]{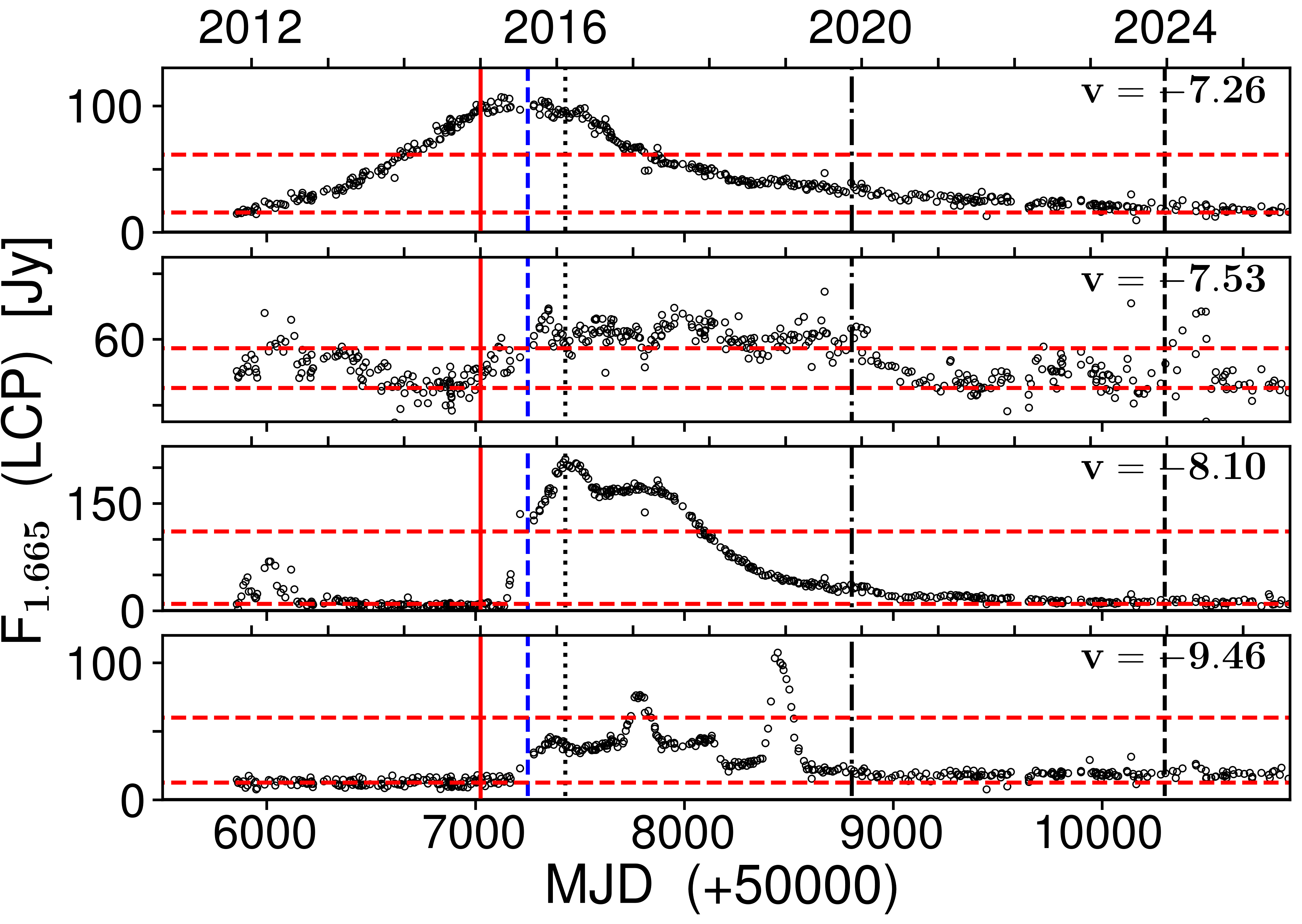}
	\caption{Time series plots of the observed flux density are presented for four 
    1.665\,GHz LCP OH velocity features. The vertical lines are as defined in Fig.\ 
    \ref{fig:IntfluxMeth}.}
 \label{fig:LCP}
\end{figure}

Eight OH velocity features are selected from Figs \ref{fig:spectra} and 
\ref{fig:ds_oh1665} with maxima occurring after the 2015 MM1 event. Time series 
plots of the flux density in four RCP and LCP velocity channels are presented in 
Fig.\ \ref{fig:RCP} and \ref{fig:LCP}, respectively. Only the bottom two RCP 
features of these OH time series profiles are similar to each other. Seven of the 
eight profiles appear to have at least two maxima, with the exception of 
$v_\textrm{1.665L} = -7.26$\kms, but at different times. The $v_\textrm{1.665L} = 
-7.53$\kms\ profile is the least pronounced. There is obvious periodic flaring in 
the $v_\textrm{1.665R} = -7.35$\kms\ feature (see Fig.\ \ref{fig:RCP}; this is a 
contaminating feature associated with G351.161+0.697). However, a line blended 
feature of it also appeared to flare in association with the 2015 MM1 flaring 
event. Excluding $v_\textrm{1.665L} = -7.53$\kms, all are either at, or nearing, 
quiescence. It is not apparent if any of these components experienced flaring 
associated with the recent 6.7\,GHz \meth\ maser flaring identified above in 
Fig.\ \ref{fig:IntfluxMeth}.

The LCP feature, $v_\textrm{1.665L} = -7.26$\kms\, has the same velocity and reaches 
its maximum before the brightest 6.7\,GHz \meth\ feature, $v_{6.7} = -7.26$\kms\ 
reported in \citet{MacLeod2018}. It experienced a slow, 0.07\,Jy\,d$^{-1}$, flux 
density increase between late 2011 to early 2015. It returned to its late 2011 value 
in 2024. The onset of this feature began more than three years prior to all the other 
identified features including the \meth\ and \water\ masers. Its RCP counterpart, 
identified in \citet{Chanapote2019} as $v_\textrm{1.665R} = -10.64$\kms, is line 
blended with an G351.161+0.697 feature (see Fig.\ \ref{fig:ds_oh1665}). Its apparent velocity 
drift, in the dynamic spectrum here, varies annually; it is not possible to determine 
its correlation with $v_\textrm{1.665L} = -7.26$\kms.

\subsubsection{Gaussian fit OH features}
OH spectra tend to be maser rich sources resulting in many line-blended velocity 
features, as can be seen in Fig.\ \ref{fig:spectra}. A single channel, or an 
integrated flux density in a specified velocity extent, may not be enough to 
identify velocity features during flaring. A measurement of variations in 
spectral shapes may be a better metric to identify such features. This is 
achieved by fitting multiple Gaussian spectral profiles.

\begin{figure*}   
	\includegraphics[clip,width=\textwidth]{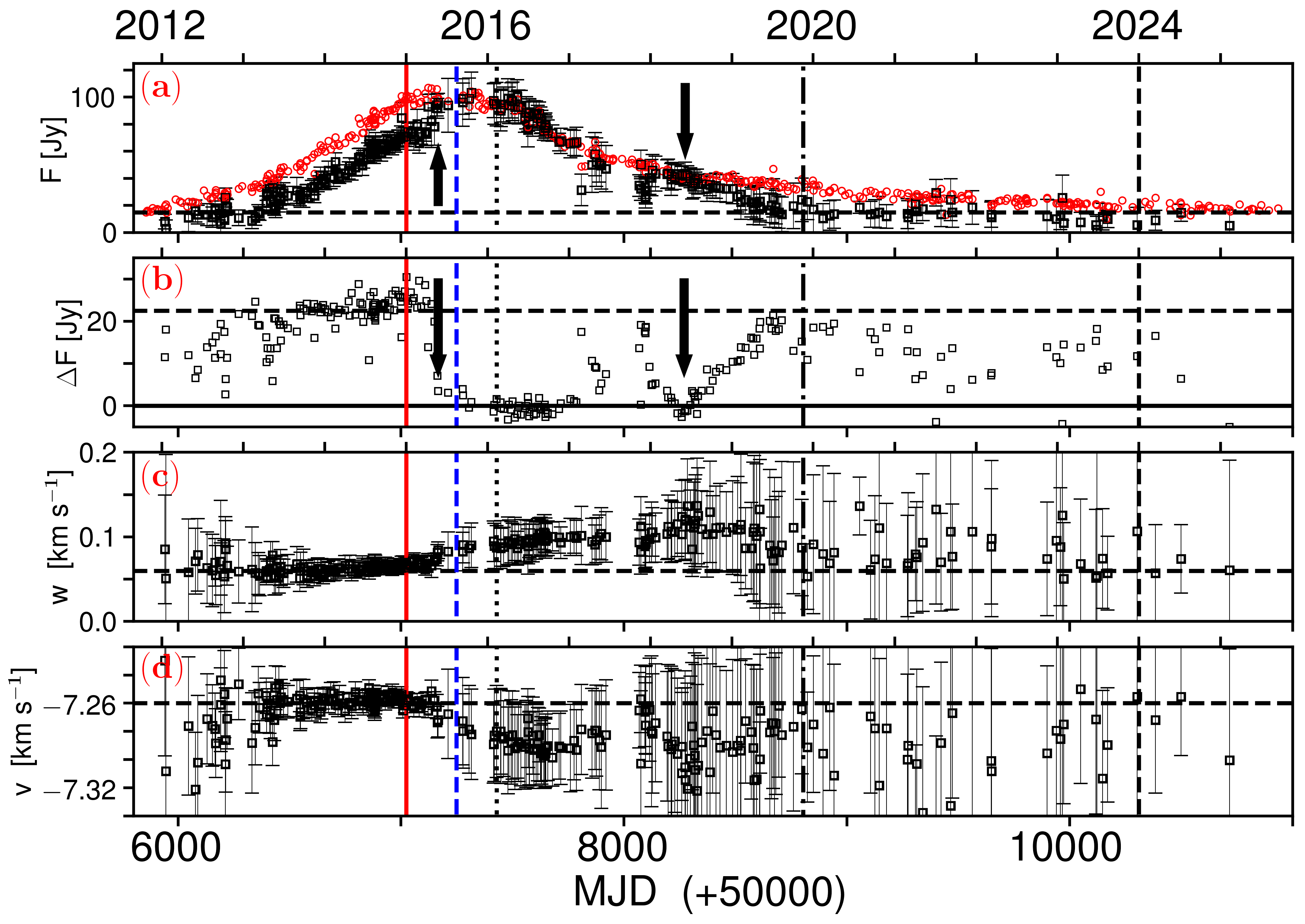}
	\caption{Time series data of the parameters of Gaussian profiles fitted to the 1.665\,GHz LCP OH velocity feature at $v_{1.665} = -7.26$\kms\ are shown as black points with error bars in panels (a), (c), and (d):
    (a) $F_\textrm{G}$ fitted shapes with $F_\textrm{C}$ velocity channel fitted shapes (red circles) included, 
    (b) difference $\Delta F = F_\textrm{C} - F_\textrm{G}$, 
    (c) $w$, and (d) $v$. 
    The vertical lines are as defined in Fig. \ref{fig:IntfluxMeth}. 
    In (a) the dashed horizontal black line denotes the quiescent level. 
    The average pre-flaring $\Delta F$ is denoted by the dashed horizontal black line in (b). 
    The pre-flaring line width value is represented by the dashed horizontal black line in (c).
    The dashed horizontal black line in (d) is the velocity channel.
    Black arrows denote flaring.}
 \label{fig:gf73}
\end{figure*}

The impact of line blending is partially mitigated by fitting multiple profiles. 
It removes flux density contributions of interfering, but visibly separate, features from those of interest. In the OH spectra presented here, e.g.\ those in Fig.\ \ref{fig:spectra}, many (10 to 20) Gaussian profiles are fitted to both polarisations in each epoch with bright ($F > 5$\,Jy) velocity features. 
The fitted shape of each feature may show sudden changes in flux density ($F_\textrm{G}$), broadening or narrowing in line width ($w$), and shifts in velocity ($v$) after the 2015 MM1 accretion event. Only one Gaussian was fitted to the $v_\textrm{1.665L} = -7.26$\kms\ feature.
The time series of these parameters are presented for $v_\textrm{1.665L} = -7.26$\kms\ in Fig.\ \ref{fig:gf73}.
For comparison, we include the LCP OH velocity channel flux density ($F_\textrm{C}$) at the same velocity.
The residual from the Gaussian model, $\Delta F = F_\textrm{C} - F_\textrm{G}$, is also plotted in Fig.\ \ref{fig:gf73}(b).

In Fig.\ \ref{fig:gf73}, the $F_\textrm{G}$ and $F_\textrm{C}$ time series show slowly brightening LCP maser emission of 0.04 and 0.07\,Jy\,d$^{-1}$ respectively, since 2011 October 21 (MJD 55856), prior to the onset of \meth\ maser flaring (in 2015). 
Note that $F_\textrm{C} > F_\textrm{G}$ by a constant value ($\sim22$\,Jy) apparent in $\Delta F$ in Fig.\ \ref{fig:gf73}(b); $F_\textrm{C}$ values may be comprised of multiple line-blended features, one constant and another (others) variable.

\begin{figure}   
	\includegraphics[clip,width=\columnwidth]{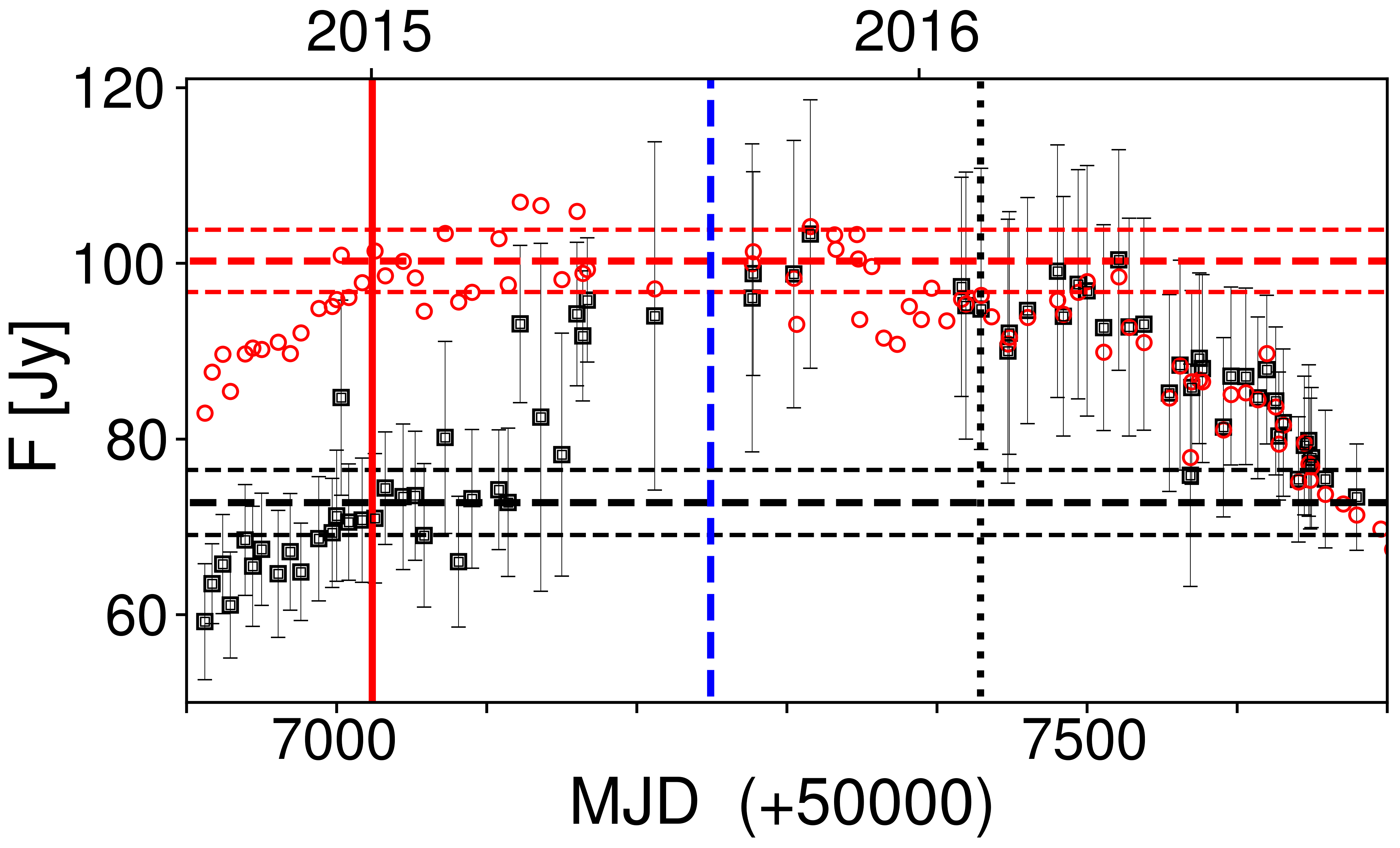}
	\caption{An expanded graphic of Fig.\ \ref{fig:gf73}(a): time series of $F_\textrm{C}$ 
    and $F_\textrm{G}$, black squares with error bars and red circles respectively, for 
    the feature $v_\textrm{1.665L} = -7.26$\kms\ are plotted. The vertical lines are as 
    defined in Fig.\ \ref{fig:IntfluxMeth}. The horizontal dashed red lines denote the 
    average velocity channel flux density (between MJD 57023 and 57348) and 1$\sigma$ 
    variance. The horizontal dashed black lines denote the average Gaussian fitted flux 
    density (between MJD 57023 and 57114) and 1$\sigma$ variance.}
 \label{fig:blowup}
\end{figure}

\begin{figure}   
	\includegraphics[clip,width=\columnwidth]{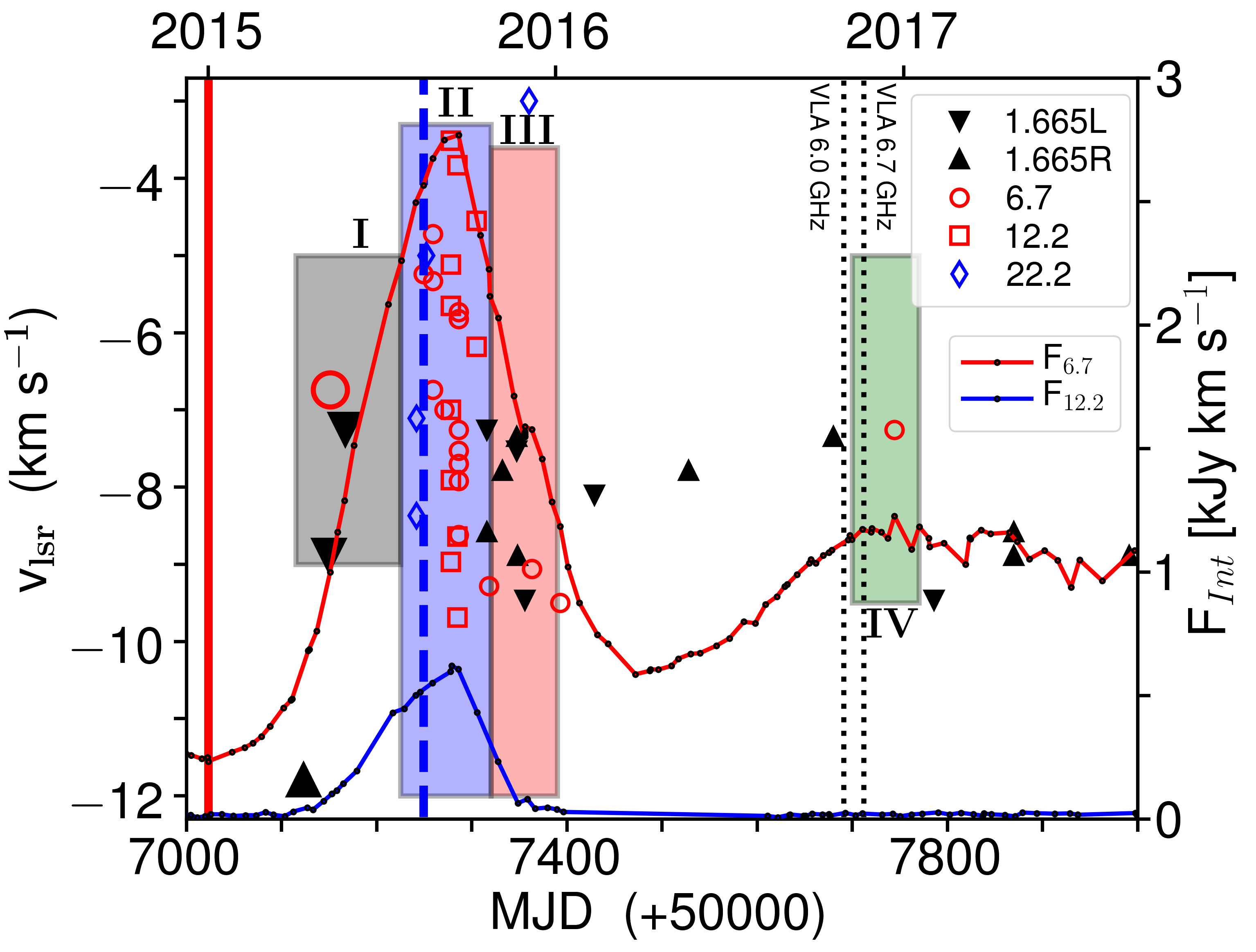}
	\caption{Plot of MJD dates where various maser features reached a maximum 
    versus their velocity. Maxima for \meth\ and \water\ maser features reported
    in \citet{MacLeod2018} are included. The 6.7 (red) and 12.2\,GHz (blue) 
    integrated flux densities (secondary y-axis) verses MJD are plotted; black 
    dots along each curve mark epoch of observation.  Enlarged symbols represent 
    the first epoch of flaring for OH and \meth\ features. Shaded Regions (I 
    through IV) denote velocity and MJD extents of maxima (colours match emboldened 
    dates in Table \ref{tab:Maxima}). Dates of onset of 6.7\,GHz flaring activity 
    and first reported maxima (in \citet{MacLeod2018}) are demarcated by solid red 
    and dashed blue vertical lines. Observation dates for 6.0\,GHz OH and 6.7\,GHz 
    \meth\ masers \citep{hunter18} are marked by dotted black vertical lines.}
 \label{fig:Max_Dates}
\end{figure}

Values of $F_\textrm{G}$, $w$, and $v$, shown in Fig. \ref{fig:gf73}(a), (b), and 
(d) respectively, determined for $v_\textrm{1.665L} = -7.26$\kms, deviated from their 
pre-flare values during the onset of flaring in MM1 (solid red line in Fig.\ 
\ref{fig:gf73}). A noticeable change in $F_\textrm{G}$ occurs weeks after the onset 
of \meth\ maser flaring, suggesting a new flaring OH feature, other than the ones 
previously varying slowly. Variations in $F_\textrm{G}$ are seen better in Fig.\ 
\ref{fig:blowup}. The transition of $v_\textrm{1.665L} = -7.26$\kms\ from a slow 
increase to a relatively constant level occurred at about the same time as the onset 
of the \meth\ and \water\ maser flaring in 2015 January 1 (MJD 57023). The onset of 
flaring of a new OH velocity feature at $v_\textrm{1.665L} \leq -7.30$\kms, denoted 
by a black arrow in Fig. \ref{fig:blowup}, beginning on 2015 April 24 (MJD 57114) 
was detectable; it rose from $73\pm4$\,Jy, the average of $F_\textrm{G}$ between MJD 
57023 to 57114, to $100\pm4$\,Jy, the average of $F_\textrm{C}$ between MJD 57023 to 
about 57350. The maxima of this flaring feature are listed in Table \ref{tab:Maxima} 
and shown in Fig.\ \ref{fig:Max_Dates} as enlarged symbols in the shaded grey region 
labelled $\textrm{I}$. These maxima occurred $\sim100$\,d before the brightest \meth\ 
maser (dashed blue line) did. This flaring feature was accompanied by line broadening 
and velocity drifting to new blue-shifted velocities. All are the result of variations 
in the spectral shape caused by at least one maser varying. 

We complete analysis, for comparison purposes to the $v_\textrm{1.665L} = -7.26$\kms\ 
feature, of the brightest MM3 1.665\,GHz OH Zeeman pair, $v_\textrm{1.665R} = 
-11.86$\kms\ and $v_\textrm{1.665L} = -8.89$\kms \citep{Chanapote2019}, in Appendix\ 
\ref{app:OHMM3}. They are slowly strengthening throughout observations, see Fig.\ 
\ref{fig:App_gf89},unlike that for $v_\textrm{1.665L} = -7.26$\kms. Results of the 
analysis of the magnetic field ($B$) are listed in Table \ref{tab:deltaB}. It was 
weakening before the 2015 MM1 event but strengthened afterwards.The change in sign 
of $\Delta B$ occurs at the intersection of the two linearly fitted lines in $B$ in 
Table \ref{tab:deltaB}, at MJD $57000\pm100$. The \meth\ maser flaring onset occurs 
within errors of this date. After MJD 59500 the variation in $B$ ceased. Regrettably, 
identification of a Zeeman counterpart to $v_\textrm{1.665L} = -7.26$\kms\ was not 
possible. No interferometric observations were made, nor could we find one correlated 
in either flux density or velocity drift here. Perhaps the RCP counterpart is weak 
and masked by other OH masers in the spectra, or the conditions in the maser cloud do 
not support it. 

\begin{table}
\centering
\caption{The parameters for the linear regression fitting to the time series for $B$ 
    for the given MJD range. The magnetic field is determined for the Zeeman pair 
    $v_\textrm{1.665L} = -8.89$\kms\ and $v_\textrm{1.665R} = -11.86$\kms\ shown in 
    Fig.\ \ref{fig:App_gf89}.} 
\label{tab:deltaB}
\setlength{\tabcolsep}{3pt} 
\begin{tabular}{cccc}
\hline
MJD  & Duration & Slope, $\Delta B$ & Intercept, $B_0$ \\
Range & [d] & [$\times 10^{-5}$\,mG\,d$^{-1}$] & [mG] \\
\hline 
55856 to 57025	& 1169 & $+$1.3$\pm$0.2 & $-$5.10$\pm$0.02 \\
57696 to 59487	& 1791 & $-$1.9$\pm$0.2 & $-$4.88$\pm$0.01\\
59487 to 60936 & 1449 & $+$0.2$\pm$0.3 & $-$5.08$\pm$0.03\\
\hline
\end{tabular}
\end{table}

Individual maxima for each time series reported here, including both velocity channel and 
Gaussian fit time series, are presented in Table \ref{tab:Maxima}. The first features to 
reach a maximum are denoted in black boldface in Table \ref{tab:Maxima} (MJD 57167). These 
are, surprisingly, about 100\,d before the maxima presented in \citet{MacLeod2018}. The 
blue boldface values in Table \ref{tab:Maxima} are the maxima of the 6.7\,GHz \meth\ 
masers in the 2015 flaring event and found in the integrated flux density time 
series here. Finally, in this table, the first maximum from the 1.665\,GHz OH features in 
Fig.\ \ref{fig:RCP} and \ref{fig:LCP} are listed in red. The results in this Table 
\ref{tab:Maxima}, including the results from \citet{MacLeod2018}, are presented in Fig.\ 
\ref{fig:Max_Dates}. We include the 6.7 and 12.2\,GHz integrated flux density time series 
from Fig.\ \ref{fig:IntfluxMeth} in this image. The cadence of observations for each are 
also included. Finally, we visually inspected each velocity channel to determine the MJD 
and velocity range in which maxima were found. These are presented as shaded rectangles 
in Fig.\ \ref{fig:Max_Dates} with colours matching those of the boldface values in Table 
\ref{tab:Maxima}. Each colour-coded area is numbered (Region I-IV).

The maxima of the \meth\ masers reported in \citet{MacLeod2018} are found here in Region 
II in Fig.\ \ref{fig:Max_Dates} while their OH maxima are in Region III. The \water\ masers 
reached their maxima much later. Here, we find the first maxima occurred in OH (enlarged 
black triangles) and 6.7 \meth\ (enlarged red circle) masers about 100\,d earlier in Region 
I. Though the cadence of observations was not uniform for each transition, apparently there 
is segregation of transition maxima over time, e.g.\ 6.7 before 12.2\,GHz \meth\ masers in 
Region II. Finally, it appears that, of the 1.665\,GHz OH features studied here, there is 
an initial flaring event (Region I) before the brightest feature at $v_\textrm{1.665L} = 
-8.10$\kms (Region III) reached its maximum reported in \citet{MacLeod2018}.

\subsection{Summary of results}

The OH masers have largely returned to quiescence since the onset of the 2015 accretion 
event in MM1, as have the 12.2\,GHz \meth\ masers. The 6.7\,GHz \meth\ and 22.2\,GHz 
\water\ masers continue to flare 10 years on. Possible 12.2\,GHz quasi-period, 
$P_\textrm{Visible} = $380$\pm$50\,d, flaring began $\sim$800\,d after the onset of the 
2015 MM1 accretion event. Seven more flaring 1.665\,GHz OH maser features associated with 
MM1 are presented here. One feature, $v_\textrm{1.665L} = -7.26$\kms, has been slowly 
strengthening since 2011 but began decaying as \meth\ maser flaring in MM1 began. The 
brightest Zeeman pair of OH masers have been, and continue to, strengthen. Three OH and 
one 6.7\,GHz \meth\ maser flared about 100\,d before the \meth\ feature at $v_{6.7} = 
-5.24$\kms\ reached its maximum reported in \citet{MacLeod2018}. A summary of analysis 
of each time series, the identification of maxima, is presented in Table \ref{tab:Maxima}.

\section{Discussion}
\label{discussion}

\subsection{A simple model}
The most studied accretion event, that in \Gthreefiveeight, has offered new insights that 
can be applied to the event in MM1. \citet{Burns2023,BurnsCorrection2023} described the 
structure of \Gthreefiveeight\ based on multi-epoch interferometric observations of masers 
associated with it. They identified spiral arms along which dust and gas accreted onto the 
proto-star's surface. Hereafter, a luminous outburst of radiation propagated outwards until 
it was absorbed by dust and gas in the surrounding medium, heating the medium. This heated 
dust and gas then re-radiated infrared photons, which were in turn absorbed and re-radiated. 
The speed of progression will likely be significantly slower than the speed of light, 
dependent upon the composition and density. 

This process produced a ``heat'' wave that propagated radially outward at the estimated speed 
$V_\textrm{G358}\sim0.04$ to $0.08\,c$ \citep{Burns20} (or $\sim$7 to 14\aud), where $c$ is 
the speed of light. \citet{Burns2023,BurnsCorrection2023} proposed the masers illuminated its 
spiral structure in the disk as the ``heat'' wave propagated outward. In such a case, velocity 
features identified in single-dish spectra may have contributing emission in multiple 
positions in the spiral structure and hence be activated at different times during this 
outward propagation. In time series, such as those provided at the Ibaraki Methanol monitoring 
programme referred to as iMet\footnote{See https://vlbi.sci.ibaraki.ac.jp/iMet/data/192.6-00/}, 
maxima are seen at different times for different velocity features in the \Gthreefiveeight\ outburst.

Based on the above model, we can assume the maxima in Fig.\ \ref{fig:Max_Dates} may denote 
structure (the shaded Regions I--IV), possibly the inner disk. The 
various maxima shown in Fig.\ \ref{fig:IntfluxMeth} (6.7\,GHz \meth\ masers) and Fig.\ 
\ref{fig:RCP} $\&$ \ref{fig:LCP} (RCP $\&$ LCP 1.665\,GHz OH masers) suggest a clumpy structure 
beyond Region IV (possibly the outer disk). Also interesting in this figure, and listed in Table 
\ref{tab:Distances}, are the apparent associations of the various maser species in each region. 
This must be related to the conditions in each region resulting in masering action. 

Regrettably, only a single epoch of interferometric observations was made for the 6.0\,GHz OH and 6.7\,GHz \meth\ masers in MM1 in 2016 November (MJD 55720) \citet{hunter18} and not for the ground state OH, nor 12.2\,GHz \meth, masers. 
\citet{csg02} present models in which OH and \meth\ masers co-exist but without knowledge of spatial associations little more can be determined.

\begin{figure}   
\includegraphics[clip,width=\columnwidth]{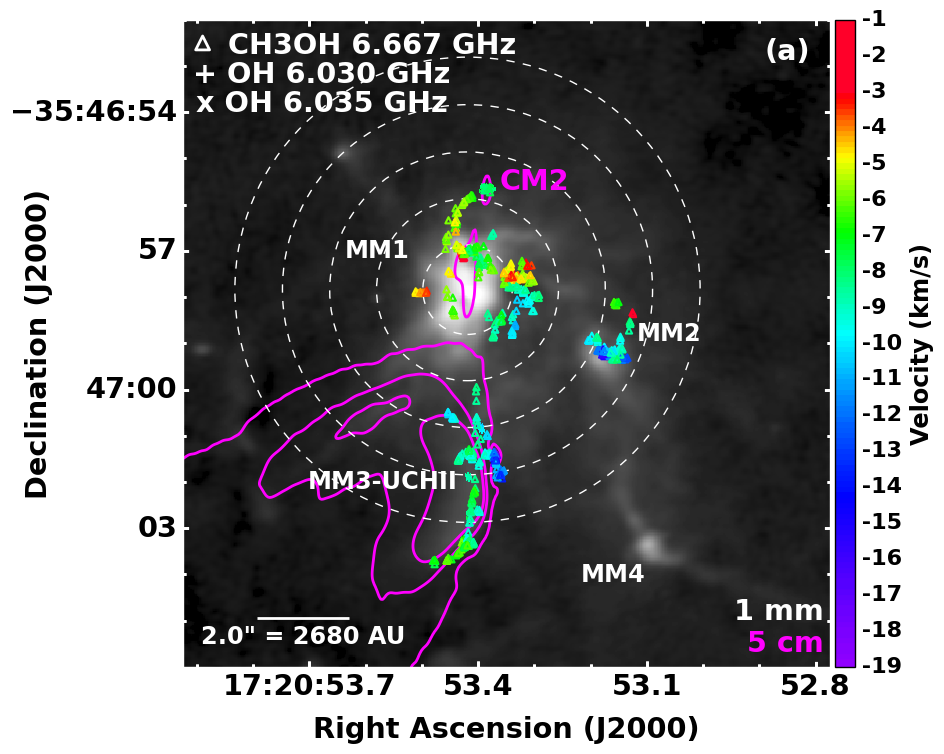}
\includegraphics[clip,width=\columnwidth]{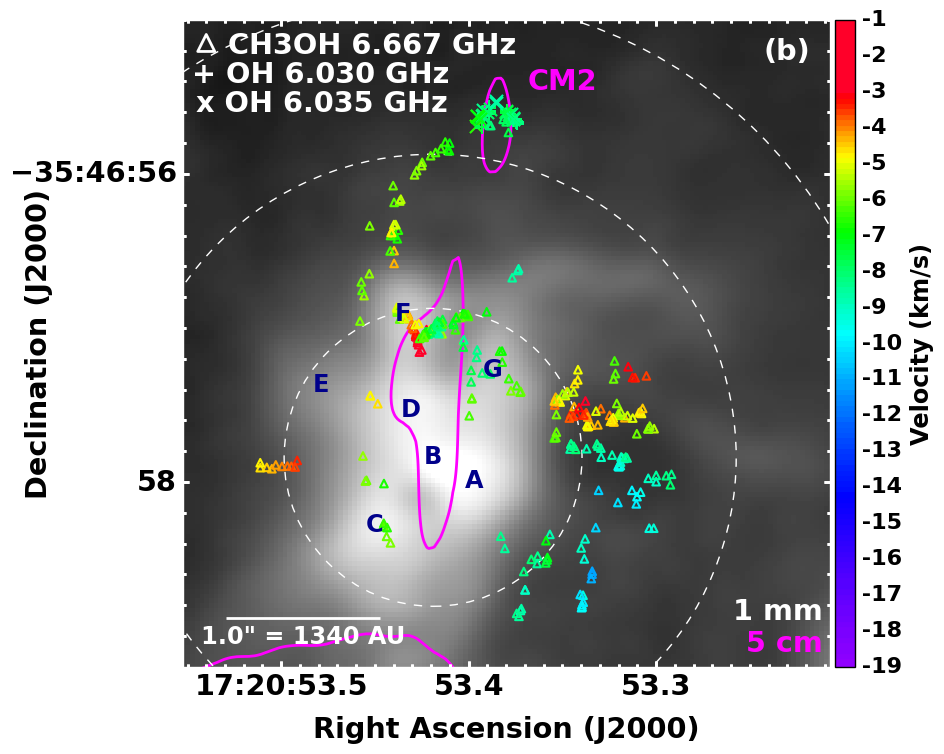}
\caption{(a) Plot of 6.0\,GHz OH and 6.7\,GHz \meth\ masers superposed on centimetric 
     continuum (contours) and millimetric continuum (greyscale) observations originally 
     shown in \citet{Brogan2016}. Concentric circles at 1\farcs0 (1340\,AU) intervals are 
     plotted centred on \ngcimm B. (b) Same as (a) but zoomed in to the region of MM1. 
     Letters A--G correspond to \meth\ maser clusters in \citet{hunter18}. The radio continuum source
     CM2 \citep{hunter18} is also present.}
 \label{fig:Spot_map}
\end{figure}

The maser spot map presented in \citet{hunter18} is presented here in Fig.\ \ref{fig:Spot_map}. 
It also includes centimetric and millimetric continuum emission from sources in \ngci. 
\citet{Reid14} determined the distance to these masers to be about 1.34\,kpc, ergo one arcsecond equals 1340\,AU.  
Concentric circles centred on MM1B, the accreting source in the 2015 event \citep{Brogan2018}, at 1\arcsec\ intervals show the association of masers to sources and projected distances. 
There are a dearth of masers inside $\sim1$\arcsec; possibly the masers in the inner disk (in Regions I--III) have been thermalized. 
Perhaps the \meth\ masers identified in MM1C and/or MM1G (see Fig.\ \ref{fig:Spot_map}) are remnants of these brightest flaring features. 
They are located about 700\,AU from MM1B and at about the distance estimated in Table \ref{tab:Distances} for Region III. 
The flaring masers in Region IV of MM1 are about 1400\,AU from MM1B, at the first concentric circle in Fig.\ 
\ref{fig:Spot_map}. 
Masers in CM2 and MM2 lie $\sim4000$\,AU from MM1B as well as the MM3 maser features at $v_\textrm{1.665L} = -7.26$\kms\ and the Zeeman pair $v_\textrm{1.665L} = -8.89$\kms\ and $v_\textrm{1.665R} = -11.86$\kms.

\subsubsection{Radially expanding ``heat'' wave}

Fortuitously, the \citet{hunter18} observations were taken when the 6.7\,GHz masers in Fig.\ \ref{fig:IntfluxMeth} reached a secondary maximum, which is $\sim675$\,d after accretion onset or in Region IV. 
Hence, the estimated speed of the radially expanding ``heat'' wave is $V_\textrm{radial} \approx 1.8$\aud. This is about 4 to 8 times slower than that reported in \Gthreefiveeight.

Estimated distances to maser regions in Fig.\ \ref{fig:Max_Dates} are presented in Table \ref{tab:Distances} assuming $V_\textrm{radial} = 1.8$\aud. 
Other masers are possibly at greater distances than the 6.7\,GHz masers in \citet{hunter18} (perhaps as far from MM1B as $\sim2700$\,AU). 

It is also possible these regions are closer to MM1B. They may be re-flaring regions resulting from thermal instabilities in the accretion disk \citep{Elbakyan2024}. 
If this were re-flaring we might expect to see renewed 12.2\,GHz \meth\ maser emission, but we report none after the initial flare and decay.

\begin{table}
\centering
\caption{Estimation of radial distance, from MM1B, to identified regions of maser activity 
assuming a radially outward expanding ``heat" wave. 
The speed of expansion is taken to be $V_\textrm{radial} = 1.8$\aud.} 
\label{tab:Distances}
\setlength{\tabcolsep}{3pt} 
\begin{tabular}{cccc}
\hline
Region & Flaring & Days & Distance\\
 & transitions & from MJD 57023 & \\
 & [GHz] & [d] & [AU] \\
\hline 
I & 1.665/6.7  & 125 & 250\\
II & 6.7/12.2/22.2 & 225 & 450\\
III & 1.665/6.7/22.2 & 325 & 650\\
IV$^{1}$ & 1.665/6.7 & 675 & 1350\\
\hline
\multicolumn{4}{l}{$^{1}$Region used to estimate expansion speed.}\\
\end{tabular}
\end{table}

\subsubsection{Along the outflow axis}
\citet{Brogan2018} reported an outflow centred on MM1B in a North-South orientation 
with one termination point at CM2, and the other superimposed on the North-Eastern 
area of MM3 (see Fig.\ \ref{fig:Spot_map}). 

The 6.7 GHz \meth\ masers show morphology consistent with tracing the edges of 
various outflow cavities \citep[compare the outflows identified in][]{Brogan2018,
Chibueze2021,Hunter2021}. The 6.7 GHz \meth\ masers are typically confined to the 
close vicinity of a HMYSO, but it is reasonable that the dense edges of the outflow 
cavities were heated to $T_\textrm{dust} \sim 400$ K due to an increase of the source's 
accretion luminosity \citep{Hunter2021,Zhu2024}.

The CM2 6.0\,GHz OH masers reached their maximum in Region II, about 225\,d after onset. 
The estimated speed along the outflow axis, using the separation between CM2 and MM1B in Fig.\ \ref{fig:Spot_map} ($\sim$4000\,AU), is $V_\textrm{outflow} \sim 18$\aud.
This is slightly faster than the speeds $V_\textrm{G358}$ and $V_\textrm{radial}$ here; possibly due to a lower density in the cavity formed by the outflow. 
\citet{caswell11} showed a magnetic field reversal in MM3 from their study of 6.0\,GHz OH masers. 
However, \citet{Brogan2018} suggest the masers with $B > 0$ are not associated with MM3 but with the MM1 outflow.

\subsection{Impact on the local environment}

\subsubsection{Simple explanation for $F_\textrm{C} > F_\textrm{G}$}

\citet{Chanapote2019} observed 1.665\,GHz OH masers towards NGC\,6334I only 10 months 
after this monitoring programme began. They reported all emission to be associated with 
MM3, while they achieved an average sensitivity level of $\sim0.1$\,Jy. They reported 
three maser spots with velocity $v_\textrm{1.665L} = -7.27$\kms; two were associated 
with the brightest OH masers, the Zeeman pair shown in Fig.\ \ref{fig:App_gf89}. The third 
was about 1\farcs7 to the South-East but will also add to the total flux density, 
$F_\textrm{Chan} = 9.10$\,Jy. 

Here we report a maser feature at $v_\textrm{1.665L} = -7.26$\kms\ but with $F_\textrm{C} 
= 26\pm1$\,Jy on 2012 August 15, at most eight days after their observations. It is 
possible that the extended emission is resolved out considering their minimum baseline 
of 100\,km. However, if any or all of the three masers vary their spectral shape may 
appear as non-Gaussian, see Appendix \ref{app:Simple_Model}. Ergo, fitting a single Gaussian 
profile may not find the same maximum $F$. Prior to flaring in MM1 in 2015, $F_\textrm{C} 
> F_\textrm{G}$ (see Fig.\ \ref{fig:gf73}). 

In Fig.\ \ref{fig:gf73} the impact of flaring is found to be present  where $\Delta F$ 
goes to zero (see arrows in the figure). The total spectrum approaches a Gaussian profile 
during the flaring. This presents an interesting possibility of disentangling strongly 
line-blended features. The temporal profiles of $F$, $w$, and $v$ during flaring can be 
used as constraints to modelling. This is left for future work.

\subsubsection{Accretion induced impact on the OH masers}

Without a multi-epoch interferometric observation programme, it is impossible to determine 
which source harbours the varying 1.665\,GHz OH masers. In this paper, we report three 
variable features that are slowly strengthening from 2011 to 2014: $v_\textrm{1.665L} = 
-7.26$\kms feature (Fig. 7, 8,and 9) and the Zeeman pair with $v_\textrm{1.665L} = 
-8.89$\kms\ and $v_\textrm{1.665R} = -11.86$\kms\ (Fig.\ \ref{fig:App_gf89}). The first 
feature ceased strengthening when the 2015 maser flaring event commenced. On the other hand, 
the third feature experienced a brief decline at this time while the second continued 
strengthening till 2026 April. The $v_\textrm{1.665L} = -7.26$\kms also shows secondary 
flaring and is apparently associated with the MM1 event. 

Prior to the 2015 event, with this information and that in \citet{Chanapote2019}, it was 
reasonable to assume these masers are associated with MM3. After the onset of accretion in 
2015, and the cessation of strengthening in some features, it is not clear where they are 
located. Possible explanations are presented in the following sections.

\paragraph{Scenario 1: A simple coincidence.}

The simplest explanation is coincidence, where the masers in MM3 may be flaring unrelated 
to events in MM1. Typically the OH and \meth\ masers have a similar velocity extent 
\citep{Breen2010} and appear spatially coincident \citep{cetal95a,c97}. The total velocity 
extent can be estimated from all \citet{hunter18} maser observations: those in MM1 
($v_\textrm{MM1} = [-11.00, -1.55]$) overlap with those in $v_\textrm{MM3} = [-18.80, 
-1.55]$\,kms. \citet{Chanapote2019} showed that, in MM3, the brightest 1.665\,GHz OH 
masers are coincident with the brightest 6.0\,GHz OH masers and, in turn, \citet{hunter18} 
found these are associated with the brightest 6.7\,GHz \meth\ masers. According to the models 
of \citet{csg02}, \meth\ and OH masers are excited in regions with similar conditions. 
In the results presented here, only three OH features were flaring prior to 2015, which are 
generally located together with the brightest MM3 \meth\ masers. The discrepant evidence 
against this scenario is the coincident onset of the 2015 accretion event with the onset of 
decay of the OH feature $v_\textrm{1.665L} = -7.26$\kms.

\paragraph{Scenario 2: Pre-heating in MM1.}

It is not clear what the energy source is for the OH masers prior to the pre-accretion 
event, if this is not a coincidence. It has been suggested that the 2015 event was an FU 
Ori-like outburst \citep{hunter17b}, and that magneto-rotational instabilities (MRI) may 
cause FU Ori-like outbursts with infrared precursors occurring even decades before the 
optical event \citep{Cleaver2023}. Alternatively, energy may be released by gravitational 
and viscous collisional processes during the infall of dust and gas onto the accretion 
disk resulting in ``pre-heating''. \citet{MacLeod2018} identified possible earlier flares 
in 1965 and 1999, which both appear to be much less energetic and with smaller maser 
variations than for the 2015 event. \citet{Elbakyan2024} proposed repeated outbursts only 
after the accretion disk is replenished. During this replenishment process disk dust may 
be heated, increasing the dust temperature and possibly energising OH molecules 
leading to maser action.

In environments with low gas and dust temperatures the OH masers will be more 
likely, and stronger than the \meth\ masers \citep{csg02}. Simulations of infrared 
emission, between 2 and 10\,$\mu$m, produced prior to an accretion event may indeed 
energise the OH masers \citep{Masley2025}. However, after the accretion outburst the 
masers near the young stellar object will be adversely affected, even terminated. For 
example, \citet{Brogan2018} found that \water\ masers in the vicinity of MM1B completely 
disappeared after the onset of accretion. Examples of pre-burst dust warming (namely in 
mid-IR light-curves) show the dusty disks are heated well before the main optical 
outburst begins \citep{Hillenbrand2018, Szegedi-Elek2020}. The magneto-rotational 
instability triggered by gravitational instability can trigger these events 
\citep{Cleaver2023} and suggest an infrared outburst preceding the optical one.

\paragraph{Scenario 3: The impact of MM1 on MM3.}

Ground state OH (1.665\,GHz) masers found superimposed on the radio continuum in 
MM3 are two to four arcseconds ($\sim2700 - 5400$\,AU) away from MM1B. The light travel 
time between MM1 and MM3 is about 15--30\,d, which is consistent with $\sim25$\,d 
observed for the brightest OH masers.  Radiation from the accretion event in MM1B may 
indeed have impacted the masers in MM3 and may have thermalized the $v_\textrm{1.665L} = 
-7.26$\kms\ feature.  This may have temporarily weakened the brightest OH masers but 
this was only detectable in the weaker of the Zeeman pair at $v_\textrm{1.665R} = 
-11.86$\kms\ (Fig. 11).  Unfortunately no other 1.665\,GHz OH maser was flaring on 2015 
January 1, the date that marked the onset of \meth\ maser flaring. No increase in the 
\meth\ maser emission plotted in Fig.\ \ref{fig:IntfluxMethMM3} appears before this date.

Previously, the masers associated with the MM1B outflow appeared superimposed on MM3 
\citep{Brogan2018}. It is possible that the MM1 outflow impinges on MM3 and that these 
masers are actually in MM3. \citet{Wolf2024} estimated the energy of the accretion 
outbursts for S255IR NIRS3 ($\sim10^{46}$\,erg), \Gthreetwothree ($\sim10^{47}$\,erg), 
\Gthreefiveeight ($\sim10^{45}$\,erg), and \ngcimm ($>10^{46}$\,erg). The estimated 
energy exerted on an area of a 1.0\,AU radius maser spot would be $\sim10^{12}$\,erg at 
the projected distance from MM1B to MM3 ($\sim4000$\,AU). This energy may be insufficient 
to disrupt and/or thermalize masers in MM3. If it was, it would further support this 
Scenario 3. 

There are arguments for and against each of these scenarios and there is no way at this time to determine which, if any, is the correct explanation.

\subsubsection{Magnetic field variations.}

\citet{MacLeod2023} reported little or no velocity drift in the Zeeman pair $v_\textrm{1.665L} 
= -8.89$\kms\ and $v_\textrm{1.665R} = -11.86$\kms. We estimated $B_\textrm{Arg} = 
-5.12\pm0.02$\,mG\ in \citet{arm00} taken 5251 days before observations in \citet{Chanapote2019} 
where $B_\textrm{Chan} = -5.03\pm0.02$\,mG were within errors of values presented here in 
Fig.\ \ref{fig:App_gf89}. There are only small variations in the magnetic field over 
$\sim10\,000$\,d. 
Our average polarisation measure is $69\pm2\%$ compared to that in \citet[$\approx$71\,$\%$,][]{arm00} and \citet[$\approx$67\,$\%$,][]{Chanapote2019}, supporting the notion that this Zeeman pair is relatively constant. 
Yet here in Fig.\ \ref{fig:App_gf89} we see weak variation above 3$\sigma$ errors. 

The amplitude of the magnetic field was weakening prior to the 2015 MM1 event but strengthened 
afterwards. After MJD 59500 the variation in $B$ ceased. Results of linear regression fitting 
in each MJD range (before, during, and after the MM1 event) are listed in Table \ref{tab:deltaB}. 
The change in sign of $\Delta B$ occurs at the intersection of the two linearly fitted lines in 
$B$ in Table \ref{tab:deltaB}, at MJD $57000\pm100$ (see Fig.\ \ref{fig:App_gf89}). The \meth\ 
maser flaring onset occurs within errors of this date. Possible variations in coincident line 
blended masers associated with $v_\textrm{1.665R} = -11.86$\kms\ may cause apparent velocity drift 
and hence apparent changes in the $B$-field. This would support either Scenario 1 or 2 above.

If the masers are not coincident, then the variations in the $B$-field may be the result 
of density variations, $B\propto n^{\kappa}$, where $n$ is the molecular density and 
$\kappa$ ranges from $\approx 0.47$ to $\approx 0.65$ \citep{Crutcher2019}. Weak shocks 
traversing through MM3 can produce density variations and the projected light travel time 
between MM1 and MM3 is about 25\,d, which is sufficient time to impact masers in MM3 and 
supports Scenario 3. Regrettably, no RCP counterpart was found for $v_\textrm{1.665L} = 
-7.26$\kms. Ergo, we are unable to find variations in the $B$-field as the accretion event 
occurred in MM1.

\subsubsection{Accretion-induced impact on the methanol masers}

\begin{figure}   
	\includegraphics[clip,width=\columnwidth]{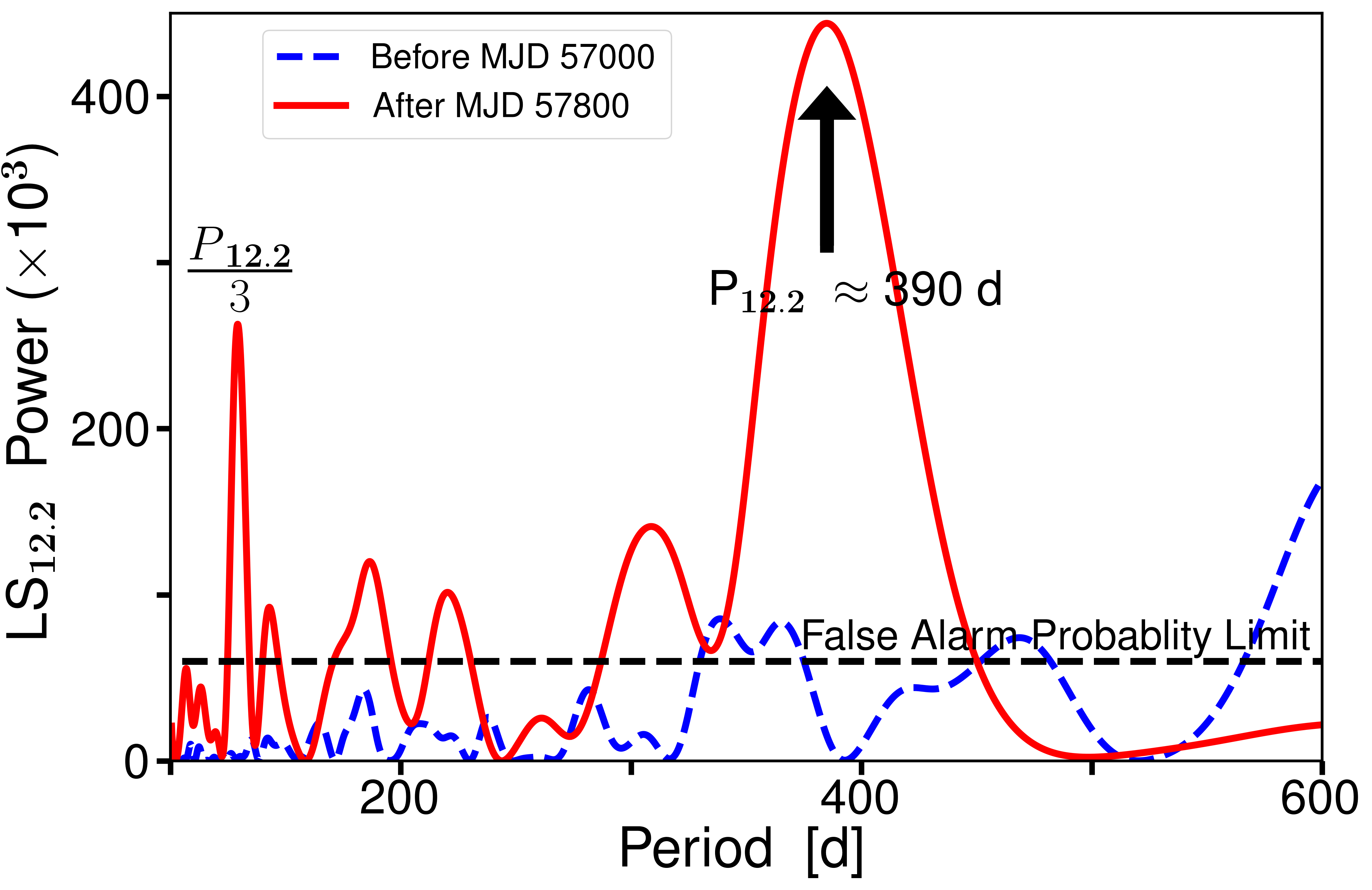}
	\caption{Analysis of 12.2\,GHz \meth\ integrated flux densities for velocity extent 
    $V_\textrm{MM2/3}$ are plotted. Dashed blue lines denote analysis of the emission 
    for all observations before MJD 57000 and solid red lines denote analysis of those 
    after MJD 57800. The average maximum value determined for a time series comprised of 
    random numbers three times the rms values determined for pre-flaring in each and run 
    1\,000 times is denoted by the horizontal black dashed lines - signals below this are not real.}
 \label{fig:LS_m12MM3}
\end{figure}

New quasi-periodic activity in the 12.2\,GHz MM2/3 dominant emission has been identified 
above in Fig.\ \ref{fig:12blowup}. This activity commenced $\sim 800$\,d after the 2015 MM1 
event and has an estimated period of $P_\textrm{Visible} = 380\pm50$\,d. Lomb-Scargle (LS) 
periodogram analysis \citep{scargle82} is applied to this data both before MJD 57000 and 
after MJD 57800. Also, false alarm probability levels are determined from the average of 
1\,000 randomly generated time series for three times the standard deviation ($3\sigma$ rms noise) in 
each dataset; below this level the signals are likely not real. The results for the 12.2\,GHz 
\meth\ maser emission in the velocity extent $V_\textrm{MM2/3}$ are plotted in Fig.\ 
\ref{fig:LS_m12MM3}. Note that a full analysis of both the 6.7 and 12.2\,GHz \meth\ maser 
emission in both velocity extents, $V_\textrm{MM1}$ and $V_\textrm{MM2/3}$, is completed, 
plotted in Fig.\ \ref{fig:App_LS_MM1and3}, and presented in Table \ref{tab:Periods}.


\begin{table}
\centering
\caption{The periods determined from the LS analysis of the 6.7 and 12.2\,GHz \meth\ 
masers in $V_\textrm{MM2/3}$ and $V_\textrm{MM1}$. Question marks indicate periods that 
may match with others but are not above the false signal level (calculated using the quoted
3$\times \sigma$ noise level).} 
\label{tab:Periods}
\setlength{\tabcolsep}{3pt} 
\begin{tabular}{lccc}
\hline
Frequency  & Velocity Extent & 3$\times \sigma$ noise& Periods \\
(GHz) & (\kms) & (Jy \kms) & (days) \\
\hline 
6.7	& [--12.0, --9.9] & 580 & 120$\pm$5?, 230$\pm$15? \\
6.7	& [--9.0, --3.0] & 105 & 115$\pm$5, 255$\pm$15 \\
12.2 & [--12.0, --9.9] & 180 & 130$\pm$5, 390$\pm$40 \\
12.2 & [--9.0, --3.0] & 15 & 400$\pm$30 \\
\hline
\end{tabular}
\end{table}

Results of the LS analysis are intriguing. There are possibly two identified periods 
in the 6.7\,GHz \meth\ emission in the velocity extent $v = [-9.0, -3.0]$\kms, (see 
Fig.\ \ref{fig:App_LS_MM1and3} and Table 
\ref{tab:Periods}). No signals are identified above the false alarm probability limit in the 
6.7\,GHz emission in the velocity extent of $V_\textrm{MM2/3}$ before or after the MM1 accretion 
event. However, one period below this limit has a value of $P_{6.7} = 230\pm15$\,d within 
errors of the strongest signal in $V_\textrm{MM1}$ but with a strength $\sim10$ times weaker. 
This is suggestive that the period occurs in the region of MM1 and not the other star-forming 
regions MM2/3. However, the flaring MM1 masers may cause false LS signals here and this 
periodicity could not be confirmed visually.

In the case of 12.2\,GHz maser emission in $V_\textrm{MM2/3}$ the periodicity is visible in Fig.\ 
\ref{fig:12blowup}. We find a LS period, $P_{12.2} = 390\pm40$\,d within errors of 
$P_\textrm{Visible}$, in the MM3-dominated velocity extent $V_\textrm{MM2/3}$. Two separate 
methods with values within errors increase the likelihood that this periodicity is real. 
A weaker signal with a similar period is detected in the MM1-dominated emission. This is 
contrary to the 6.7\,GHz emission. A secondary period, $P_{12.2} = 130\pm5$\,d, is only 
detected in $V_\textrm{MM2/3}$ and appears to be a factor of three of $P_{12.2}$, and 
hence a possible sub-harmonic. An example of harmonic signals in LS analysis are also 
seen in the $v_{6.7} = +8.8$\kms\ feature associated with G9.62+0.20 \citep{MacLeod2022}.  

It is not certain if the 12.2\,GHz periodicity was triggered by the MM1 event or whether 
it is in MM2 or MM3. A ``heat'' wave could travel $\sim 1600$\,AU in this time assuming 
$V_\textrm{radial}\approx 1.8$\aud, while at $V_\textrm{outflow} \approx 18$\aud\ the heat 
wave would travel ten times further. The implication is at the lower speed MM2 can be reached 
in this time, while at the higher speed both MM2 and MM3 can be reached. \citet{Wolf2024} 
proposed an accretion event induced periodicity in the accreting source \Gthreetwothree. 
However there is no way to determine if periodicity was present before the event. Monitoring 
leading to the detection of periodicity \citep{Proven19,MacLeod2021B} was begun $\geq$1000\,d 
after the event. There is no apparent connection between MM1 and MM2/3 save for the possible 
interaction with the outflow source associated with MM1B. Multiple epochs of interferometric 
observations at multiple maser transitions are required to confirm these results and to 
determine which sources are periodic. If such observations would confirm the periodic features 
are in MM2/3, then this may provide support for Scenario 3 above -- the accretion event in MM1 
impacted MM2/3.

\section{Summary and Future Work}
Flaring continues in the masers associated with MM1 ten years since its onset on 2015 
January 1 (MJD 57023). Methanol and OH masers are nearing prior quiescent values, 
while \water\ maser emission remains elevated. Several new flaring 1.665\,GHz OH velocity 
features associated with the MM1 accretion event are presented here. The radial infrared 
expansion speed during the accretion event is estimated to be $\sim 1.8$\aud. However, 
the expansion speed along the associated North-South outflow is about ten times faster 
($\sim18$\aud). It is proposed this expanding ``heat" wave illuminated a clumpy structure 
in MM1.

Quasi-periodic flaring, $P_{12.2} = 390\pm40$\,d, in the 12.2\,GHz \meth\ masers in 
the velocity extent dominated by MM2/3 emission is detected after the accretion event 
in MM1. Also some OH masers may have been flaring prior to the 2015 event. Three 
scenarios are presented to explain these and the impact on the local environment around 
MM1 and possibly MM2/3. Further, in particular interferometric, observations are 
required to confirm these results. 

The ground state OH 1.665\,GHz $v_\textrm{L} = -11.86$ and $v_\textrm{R}= -8.89$\kms\ lines 
are assumed to be a Zeeman pair. Magnetic fields derived from $\Delta V_\textrm{Z}$ indicate 
an increasing $B$ field before the flare, that changed to a decreasing field after the flare 
and then to a nearly constant value from MJD 59487 onwards.

\section*{Acknowledgments}
G.C.M. would like to thank I. \& D. MacLeod and Clan Machine Solutions Inc.\ for their financial 
support. G.C.M. also acknowledges financial support from the Chinese Academy of Sciences 
President's International Fellowship Initiative under grant no. 2025PVA0103 and the National 
Key R\&D Program of China  under grant no. 2022YFA1603103. JV acknowledges support from the 
Academy of Finland grant No 348342. A.C.G. acknowledges support from PRIN-MUR 2022 20228JPA3A 
``The path to star and planet formation in the JWST era (PATH)'' funded by NextGeneration EU 
and by INAF-GoG 2022 ``NIR-dark Accretion Outbursts in Massive Young stellar objects (NAOMY)'' 
and Large Grant INAF-2024 ``Spectral Key features of Young stellar objects: Wind-Accretion LinKs 
Explored in the infraRed (SKYWALKER)''.

\section*{Data availability}
The data underlying this article were accessed from the \mbox{HartRAO} data archive. 
The derived data generated in this research will be shared upon reasonable request to 
G. MacLeod.

\bibliographystyle{mnras}
\bibliography{NGC6334I_Preheating} 

\appendix 

\section{Analysis of the brightest OH masers in MM3}
\label{app:OHMM3}
\begin{figure*}   
	\includegraphics[clip,width=\textwidth]{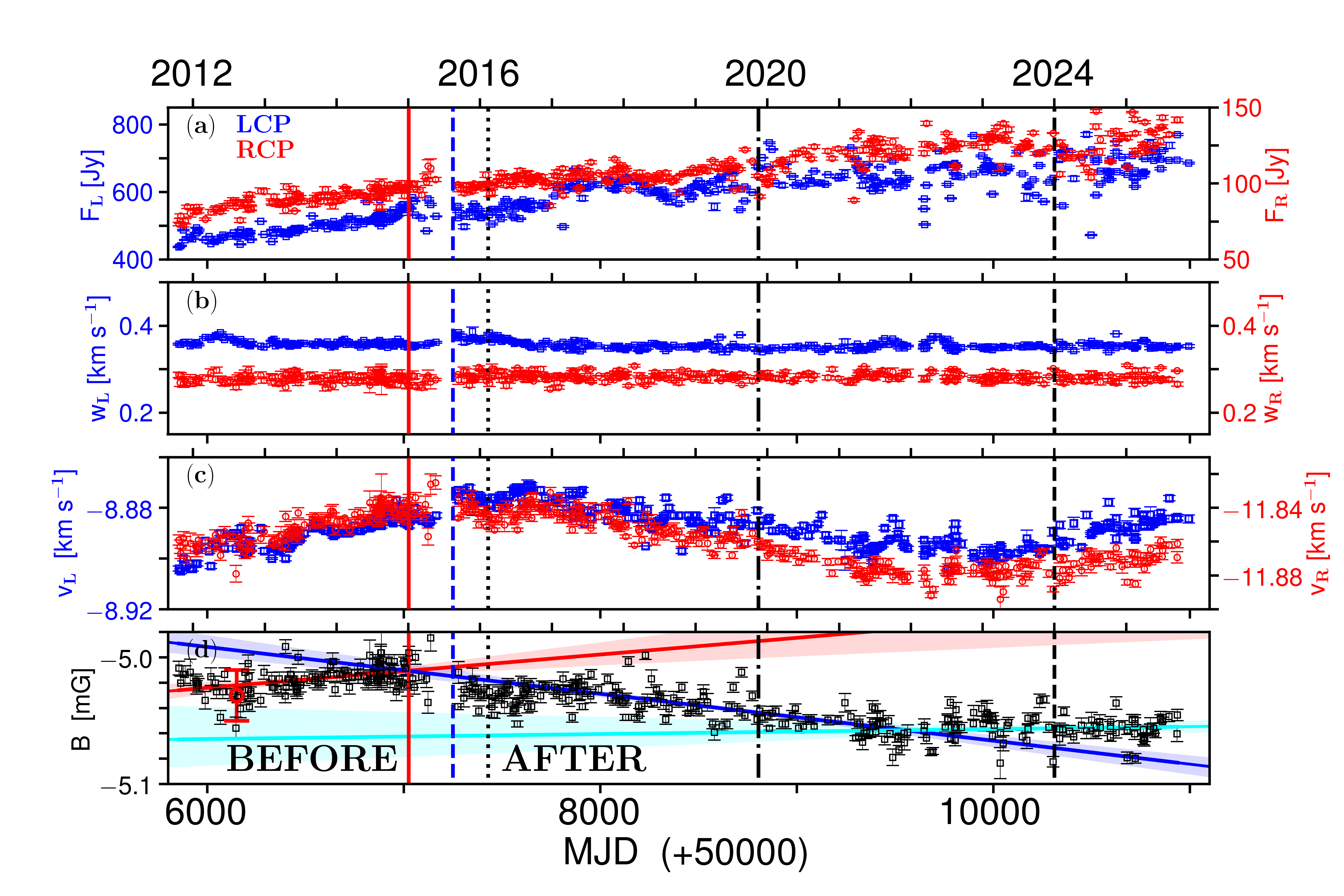}
	\caption{Time series plot of fitted Gaussian profile to the 1.665\,GHz OH Zeeman 
    pair for velocity feature $v_\textrm{1.665L} = -8.89$\kms\ (blue) and $v_\textrm{1.665R} 
    =-11.86$\kms\ (red) are presented: (a) flux density $F_\textrm{G}$, (b) line width $w$, 
    (c) velocity $v$, and (d) magnetic field $B$. The vertical lines are as defined in Fig.\ 
    \ref{fig:IntfluxMeth}. Linear regression fitting is applied to three MJD ranges (see 
    Table \ref{tab:deltaB}) to demonstrate variations in $B$. The red circle symbol marks 
    the value from \citet{Chanapote2019}.}
 \label{fig:App_gf89}
\end{figure*}

Results for fitting Gaussian profiles to the brightest MM3 Zeeman pair, $v_\textrm{1.665R} = -11.86$\kms\ 
and $v_\textrm{1.665L} = -8.89$\kms \citep{Chanapote2019}, are shown in Fig.\ \ref{fig:App_gf89}.
The flux densities and velocities of the Zeeman 
pair counterparts are relatively well correlated. Their Pearson coefficients are 0.84 and 0.65 
respectively (with no correlation in line width), supporting the argument
they are a Zeeman pair. Each continue 
to increase, unlike the $v_\textrm{1.665L} = -7.26$\kms\ feature.

The LSR velocity correction applied at 
HartRAO does not include the gravitational ``wobble'' resulting from planetary motion, 
in particular, Jupiter ($\sim12.46$\ms), Saturn ($\sim2.75$\ms) and the variance of Earth's 
eccentricity ($\sim1.5$\ms). The average error in the velocities fitted are $\sim1.8$ 
and 2.7\ms\ for the features at --8.89 and --11.86\kms\ respectively; smaller than 
the ``wobble'' resulting from the planets and that of the average error for 
$v_\textrm{1.665L} = -7.26$\kms\ ($\sim34$\ms). The effects of the planets are seen 
in Fig.\ \ref{fig:App_gf89} but not Fig.\ \ref{fig:gf73} as a result of this. These errors
are removed by determining $v_\textrm{RCP} - v_\textrm{LCP}$; when divided by the Zeeman 
splitting coefficient ($Z = 0.59$\kms\,mG$^{-1}$) we determine magnetic fields ($B$), 
see Fig.\ \ref{fig:App_gf89}(d). 

Linear regression fitting is completed in each MJD range before, during, and after 
the MM1 flaring event. Results are listed in Table \ref{tab:deltaB} and plotted, 
with 98 percentile errors, in Fig.\ \ref{fig:App_gf89}(d). \citet{Chanapote2019} 
report the velocities of the Zeeman pair and $B_\textrm{Chan} = -5.0\pm0.2$\,mG; the 
error is overestimated using their observation's velocity resolution. Their magnetic 
field employing the method here is $B_\textrm{Chan} = -5.03\pm$0.02\,mG assuming the 
velocity error is about 0.01\kms; this, too, is plotted in Fig.\ \ref{fig:App_gf89}.

\section{Simple model of Gaussian fitting}
\label{app:Simple_Model}

\begin{figure}   
	\includegraphics[clip,width=\columnwidth]{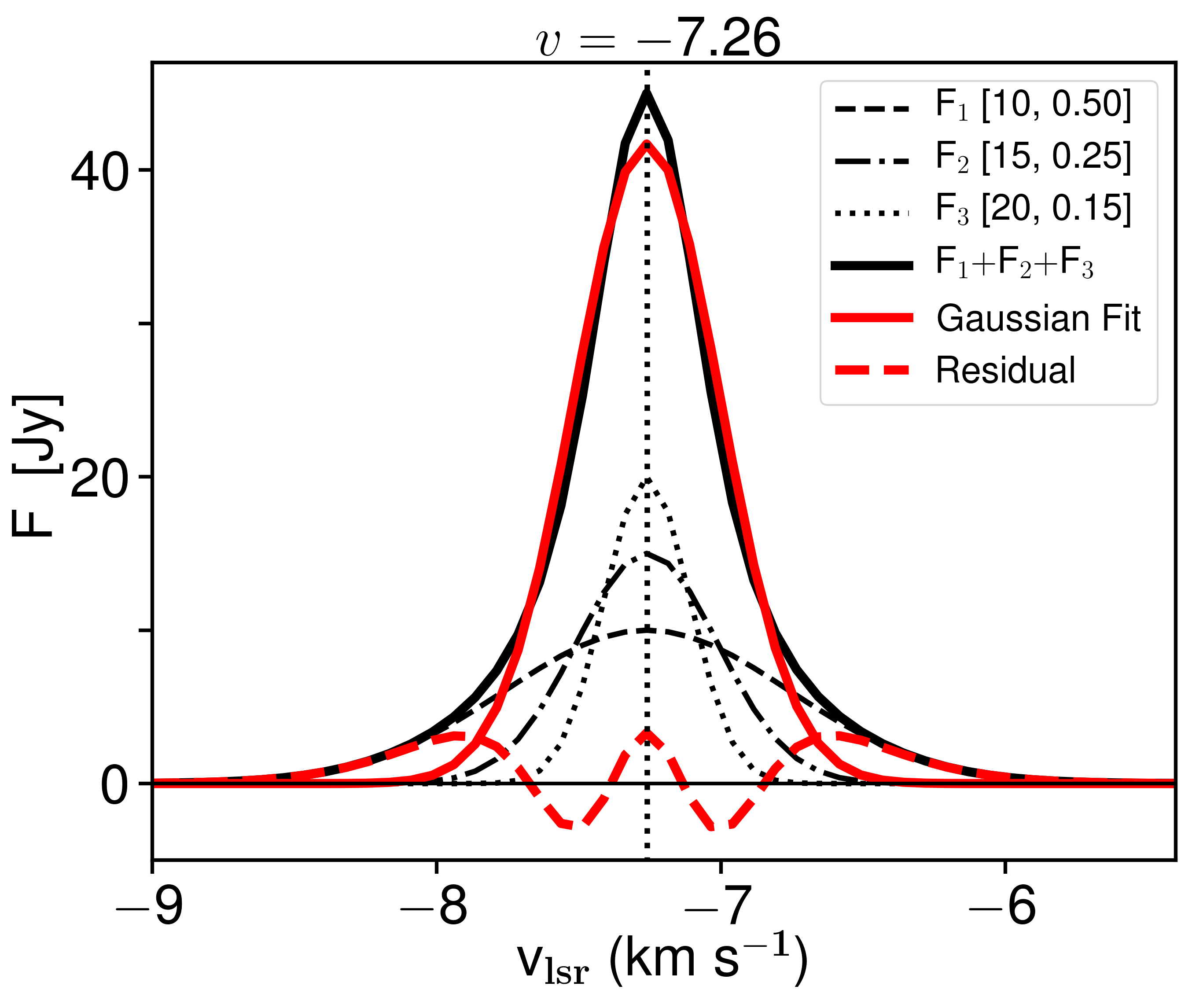}
	\caption{Plot of three Gaussian profiles, each with $v = -7.26$\kms\ but different
    flux densities and line widths (given in [$F$,$w$]), their added spectrum, a single 
    Gaussian fit, and residual spectra.}
 \label{fig:App_Gaussian_model}
\end{figure}

A simple model is constructed to explain why $F_\textrm{C} \neq F_\textrm{G}$.
Three masers with the same velocity, $v_\textrm{lsr} = -7.26$\kms\, but different $F$ and $w$, 
and the total spectrum ($F_{1}+F_{2}+F_{3}$) are plotted in Fig.\ \ref{fig:App_Gaussian_model}. 
A single Gaussian profile is 
fitted to the total spectrum and the residual $\Delta F = F_\textrm{C} - F_\textrm{G} \neq 
0$\, Jy; the total spectrum is clearly non-Gaussian. 

\section{Complete Lomb-Scargle Analysis}
\label{app:LS}
\begin{figure*}   
	\includegraphics[clip,width=\textwidth]{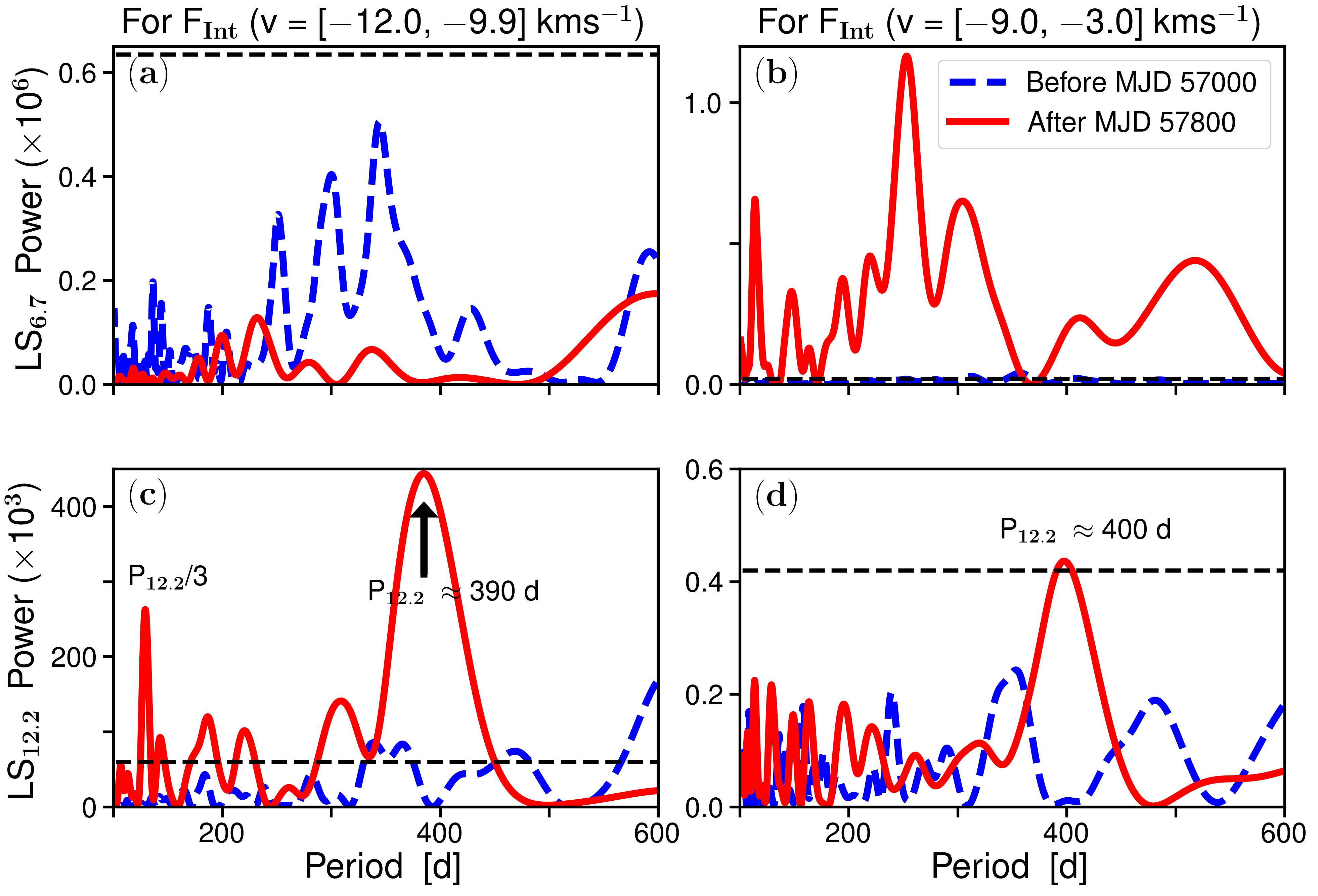}
	\caption{Analysis of 6.7 (a and b) $\&$ 12.2\,GHz \meth\ (c and d) integrated flux 
    densities are plotted. The plots in (a) and (c) represent the integrated emission 
    in the velocity extent $V_\textrm{MM2/3}$ and (b) and (d) represent that for the 
    velocity extent $V_\textrm{MM1}$. Dashed blue lines denote analysis of the emission 
    for all data before MJD 57000 and solid red lines denote analysis of of the all data 
    after MJD 57800. The 
    average maximum value determined for a time series comprised of random numbers three 
    times the rms values determined for pre-flaring in each and run 1\,000 times is 
    denoted by the horizontal black dashed lines - signals below this are not real.}
 \label{fig:App_LS_MM1and3}
\end{figure*}

Results of Lomb-Scargle Analysis for both the 6.7 and 12.2\,GHz \meth\ maser emission
in the two velocity extents $V_\textrm{MM1}$ and $V_\textrm{MM2/3}$ are shown in Fig.\ \ref{fig:App_LS_MM1and3}.

\bsp	
\label{lastpage}
\end{document}